\documentclass{ptephy_v1} 

\preprintnumber{XXXX-XXXX} 

\usepackage{amsmath}
\usepackage{graphics}
\usepackage{multirow}
\usepackage{url}

\usepackage{color}

\begin{document}

\title{Gamma Ray Spectrum from Thermal Neutron Capture on Gadolinium-157}


\author[1]{Kaito Hagiwara}
\author[2]{Takatomi Yano}
\author[1]{Tomoyuki Tanaka}
\author[1]{Pretam Kumar Das}
\author[1,4,*]{Sebastian Lorenz}
\author[1]{Iwa Ou}
\author[1]{Takashi Sudo}
\author[1]{Mandeep Singh Reen}
\author[1]{Yoshiyuki Yamada}
\author[1]{Takaaki Mori}
\author[1]{Tsubasa Kayano}
\author[1,5]{Rohit Dir}
\author[1]{Yusuke Koshio}
\author[1,*]{Makoto Sakuda}
\author[3]{Atsushi Kimura}
\author[3]{Shoji Nakamura}
\author[3]{Nobuyuki Iwamoto}
\author[3]{Hideo Harada}
\author[4]{Michael Wurm}
\author[6]{William Focillon}
\author[6]{Michel Gonin}
\author[1,7,*]{Ajmi Ali}
\author[7]{Gianmaria Collazuol}
\affil[1]{Department of Physics, Okayama University, Okayama 700-8530, Japan }
\affil[2]{Department of Physics, Kobe University, Kobe, Hyogo 657-8501, Japan}
\affil[3]{Japan Atomic Energy Agency, 2-4 Shirakata Shirane, Tokai, Naka, Ibaraki 319-1195, Japan}
\affil[4]{Institut f\"ur Physik, Johannes Gutenberg-Universit\"at Mainz, 55128 Mainz, Germany}
\affil[5]{Present address: Research Institute $\&$ Department of Physics and Nano Technology, SRM 
University, Kattankulathur-603203, Tamil Nadu, India}
\affil[6]{D\'epartement de Physique, \'Ecole Polytechnique, 91128 Palaiseau Cedex, France}
\affil[7]{University of Padova and INFN, Italy}
\affil[ ]{\email{slorenz@uni-mainz.de, aliajmi@okayama-u.ac.jp, sakuda-m@okayama-u.ac.jp}}

\begin{abstract}%
We have measured the $\gamma$-ray energy spectrum from the  thermal neutron capture, 
${}^{157}$Gd$(n,\gamma){}^{158}$Gd, on an enriched $^{157}$Gd target (Gd$_{2}$O$_{3}$) in the 
energy range from 0.11 MeV up to about 8 MeV. The target was placed inside the germanium 
spectrometer of the ANNRI detector at J-PARC and exposed to a neutron beam from the Japan 
 Spallation Neutron Source (JSNS). Radioactive sources ($^{60}$Co, $^{137}$Cs, and $^{152}$Eu) and the 
reaction $^{35}$Cl($n$,$\gamma$) were used to determine the spectrometer's detection efficiency for 
$\gamma$ rays at energies from 0.3 to 8.5 MeV. Using a Geant4-based Monte Carlo simulation of the 
detector and based on our data, we have developed a model to describe the $\gamma$-ray spectrum from 
the thermal ${}^{157}$Gd($n$,$\gamma$) reaction. 
While we include the strength information of 15 prominent peaks above 5 
MeV and associated  peaks below 1.6 MeV from our data directly into the model, we rely on
the theoretical inputs of nuclear level density and the photon 
strength function of ${}^{158}$Gd to describe the continuum $\gamma$-ray spectrum from the  ${}^{157}$Gd($n$,$\gamma$) reaction. Our model combines these two components. 
The results of the comparison between the observed $\gamma$-ray spectra from the reaction and the model are reported in detail.
\end{abstract}
\subjectindex{D21 Models of nuclear reactions, F22 Neutrinos from supernova and other astronomical objects, C43 Underground experiments, F20 Instrumentation and technique, H20 Instrumentation for underground experiments, H43 Software architectures
}

\maketitle

\section{Introduction}

Gadolinium, ${}^{\mspace{7mu}A}_{64}$Gd, is a rare earth element. Its natural composition 
(${}^{\mathrm{nat}}$Gd) includes isotopes with the atomic mass numbers $A=$ 152,154-158 and 160. 
The element features the largest capture cross-section for thermal neutrons among all stable 
elements: $\sim49000$ b. This is due to the contributions of the isotopes ${}^{155}$Gd 
(60900 b~\cite{Mughabghab2006:NeutRes}) and especially ${}^{157}$Gd 
(254000 b~\cite{Mughabghab2006:NeutRes}). 

In nuclear physics, gadolinium isotopes have been studied in neutron capture $\gamma$-ray 
spectroscopy and photoabsorption measurements to obtain information on its nuclear 
structure and properties~\cite{Groshev1962:GdGam, Bollinger1970:NeutARC, Vasilev1971:E1Params,
Voignier1986:GamSpecFromNeutInt,  Greenwood1987:ARC, Kopecky1993:Gd157, Sakurai2002:GdNCT, 
Leinweber2006:GdXSec, Chyzh2011:DANCE, Baramsai2013:DANCE, Kroll2013:DANCE, Choi2014:GdXSec, 
Valenta2015:GdTwoStepGamCasc}. The spectroscopic $(n,\gamma)$ measurements allow to catalogue neutron 
capture resonances and to probe the high density of nuclear energy levels around the 
neutron separation energy $S_n$ in the product nucleus ${}^{A+1}$Gd. Moreover, they allow to 
identify 
discrete nuclear states between the ground state and $S_n$ of ${}^{A+1}$Gd. 
Together with the inverse reaction $(\gamma,n)$ in photoabsorption measurements, the neutron 
capture $\gamma$-ray spectroscopy allows to determine the nuclear level density and the photon 
strength function of Gd. 

Recently, natural gadolinium also plays a role in experimental neutrino physics
through the identification of the electron anti-neutrino ($\overline{\nu}_{e}$) interactions.
The presence of gadolinium
enhances the 
tagging of the neutron produced from the inverse beta decay (IBD) reaction of the MeV $\bar\nu_e$ on a 
free proton: $\overline{\nu}_{e} + p \rightarrow n + e^+$. Until now, the element has been used 
as neutron absorber only by 
scintillator-based detectors~\cite{Abe2012:DC_NuEbarDisapp, An2012:DB_NuEbarDisapp, 
Ahn2012:RENO_NuEbarDisapp, Oguri2014:PANDA}. However, the addition of gadolinium 
to water Cherenkov neutrino detectors is studied and will soon be applied on a large scale in 
Super-Kamiokande (SK)~\cite{Sekiya2016:SkGd, Watanabe2009:NeutTagWCGd}. 

One important property of the 
${}^{A}$Gd$(n,\gamma)$ reaction is that the deexcitation of the compound 
nucleus ${}^{A+1}$Gd${}^{*}$ proceeds not necessarily by one but by a cascade of on average four 
$\gamma$-ray emissions \cite{Chyzh2011:DANCE}. Due to the Cherenkov 
threshold, the variable number of $\gamma$ rays and 
their energy distributions within the cascades effectively decreases the mean visible energy 
release from the neutron capture to below the Q-value. As a consequence, a reliable assessment of 
neutron tagging efficiencies in Cherenkov detectors with the help of Monte Carlo (MC) simulations 
strongly depends on a precise model for the full $\gamma$-ray energy spectrum from the thermal 
Gd$(n,\gamma)$ reaction. More seriously, such a model is important for 
non-hermetic $\overline{\nu}_{e}$ monitors~\cite{Oguri2014:PANDA}, where 
an accurate assessment of their neutron detection efficiency strongly depends on a precise model for the 
$\gamma$-ray energy spectrum from Gd$(n,\gamma)$.

There have been several publications on measured $\gamma$-ray spectra from Gd($n,\gamma$) 
reactions for neutron energies ranging from meV to MeV~\cite{Groshev1962:GdGam, 
Bollinger1970:NeutARC, Voignier1986:GamSpecFromNeutInt, Ali1994:Gd157, Valenta2015:GdTwoStepGamCasc}. 
Recently, the Detector for Advanced Neutron Capture Experiments (DANCE) at the Los Alamos Neutron 
Science Center (LANSCE) has extensively studied the $\gamma$-ray energy spectra from the radiative 
neutron capture reaction at various multiplicities in the neutron kinetic
energy range from 1 to 300 eV for ${}^{152,154,155,156,157,158}$Gd 
targets~\cite{Chyzh2011:DANCE, Baramsai2013:DANCE, Kroll2013:DANCE}. Their comparison of the data 
to MC simulations with the DICEBOX package~\cite{Becvar1998:DICEBOX} showed fair 
agreement. 
There are some publications~\cite{Groshev1959:GdGam, Ali1994:Gd157} measuring 
prompt prominent $\gamma$ rays with limited acceptance, 
but there have been  few measurements of the prompt $\gamma$ rays covering 
almost the full spectrum from 0.1 MeV to 9 MeV  from the capture reaction on ${}^{157}$Gd 
at thermal neutron energies, which enable us to compare them with 
 the modeling in Monte Carlo simulation. 

In the following, we report on a measurement of the $\gamma$-ray energy spectrum from the 
radiative thermal neutron capture on an enriched ${}^{157}$Gd sample with excellent 
$\gamma$-ray energy resolution, high statistics and low background. It was performed with the 
germanium (Ge) spectrometer of the Accurate Neutron-Nucleus Reaction Measurement Instrument 
(ANNRI)~\cite{Igashira2009:MLF-BL04, Kin2011:ANNRI, Kino2011:ANNRI, Kimura2012:ANNRI, 
Kino2014:ANNRI} that was driven by a pulsed neutron beam from the Japan 
Spallation Neutron Source (JSNS) at the Material and Life Science Experimental Facility (MLF) of 
the Japan 
Proton Accelerator Research Complex (J-PARC)~\cite{Nagamiya2012:JPARC}. Using the time-of-flight 
(TOF) method, capture reactions of neutrons in the energy range from 4 to 100 meV could be 
accurately selected for the analysis. The obtained data 
covers the entire spectrum 
from 0.11 MeV to about 8 MeV with observed 
 $\gamma$-ray multiplicities one to three. 
 Based on our data 
and a Geant4~\cite{Agostinelli2003:Geant4, Allison2006:Geant4} detector simulation of our setup, we 
have developed a model to generate the full $\gamma$-ray spectrum from the thermal 
${}^{157}$Gd($n$,$\gamma$) reaction. 
This constitutes an important step towards a corresponding model for the 
${}^{\mathrm{nat}}$Gd($n$,$\gamma$) reaction, which is ultimately relevant for neutrino detectors 
with gadolinium loading.


\section{Physics Motivation}
\label{sec:Motiv}

It is a common technique for $\overline{\nu}_{e}$ detection in the MeV regime to search for the 
delayed coincidence signals from the products of the IBD reaction $\overline{\nu}_{e} + p 
\rightarrow n + e^+$, which has a threshold energy of about 
1.8 MeV~\cite{Reines1953:NeutDet1, Reines1956:NeutDet2}:

The ``prompt signal'' occurs a few nanoseconds after the interaction and originates from the energy 
loss and the annihilation of the emitted positron. At low energies, when the invisible recoil 
energy of the neutron can be neglected, one can reconstruct the $\overline{\nu}_{e}$ energy from 
the prompt event's visible energy $E_{\mathrm{prompt}}$ as $E_{\overline{\nu}} = 
E_{\mathrm{prompt}} + 0.782$ MeV~\cite{Bellini2010:BxGeoNeut}. 

The ``delayed signal'' stems from the $\gamma$-ray emission following the capture of the 
thermalized neutron on a nucleus of the detector's neutrino target material. Neutrons produced by 
neutrinos in the MeV regime via the IBD reaction typically have kinetic energies up to several tens 
of keV and interact between ten to twenty times via elastic scattering with hydrogen before they 
are thermalized~\cite{Amaldi1935:SlowNCap, Fermi1950:Book}. The mean timescale 
$\tau_{\mathrm{cap}}$ for the neutron capture depends on the concentrations $n_i$ and the thermal 
neutron capture cross-sections $\sigma_{\mathrm{cap},i}$ of the nuclei $i$  in the detector 
material as well as on the mean velocity $v_n$ of the produced neutrons: 
$\tau_{\mathrm{cap}} \propto 1/(n_i \, \sigma_{\mathrm{cap},i} \, v_n)$. With 
hydrogen, carbon and 
oxygen nuclei naturally being present in common low-energy $\overline{\nu}_{e}$ detectors, e.g., 
organic liquid scintillator and water Cherenkov detectors, the mean neutron capture time is usually 
on the order of a few tens to hundreds of microseconds. Table~\ref{tbl:CommonCatcherIsotopes} 
summarizes thermal neutron capture cross-sections and Q-values for the most abundant isotopes of 
these elements.

\begin{table}[b!t]
    \centering
    \caption{Cross-sections~\cite{Mughabghab2006:NeutRes} and Q-values~\cite{Choi2007:EGAF} for 
             radiative thermal neutron capture reactions on nuclei naturally present in organic 
             liquid scintillator and water Cherenkov detectors.}
    \label{tbl:CommonCatcherIsotopes}
    \begin{tabular}{rcc}
        \hline
        Isotope & Cross-section & Q-value \\
                &  [mb]         & [MeV]   \\
        \hline
        &&\\[\dimexpr-\normalbaselineskip+0.1em]
        ${}^{1}$H  & 332.6  & 2.2 \\[0.1em]
        ${}^{12}$C & 3.53   & 4.9 \\[0.1em]
        ${}^{16}$O & 0.190  & 4.1 \\[0.1em]
     {${}^{157}$Gd} &   {2.54$\times$10$^8$ } &   {7.9} \\[0.1em]
       \hline
    \end{tabular}
\end{table}

  Recently, it has become a common technique to add a mass fraction of 0.1-0.2\% of gadolinium
into the neutrino targets of organic liquid scintillator~\cite{Abe2012:DC_NuEbarDisapp, 
An2012:DB_NuEbarDisapp, Ahn2012:RENO_NuEbarDisapp} and water Cherenkov~\cite{Beacom2004:GdinWCdet, 
Dazeley2009:NeutDetWCGd, Watanabe2009:NeutTagWCGd, Sekiya2016:SkGd} detectors in order to 
enhance the neutron tagging efficiency for IBD events. This basic technique was first demonstrated 
in the discovery of the neutrinos with a cadmium-loaded liquid scintillator in 
1956~\cite{Reines1953:NeutDet1, Reines1956:NeutDet2}. On the multi-kiloton scale, 
$\overline{\nu}_{e}$ detection with gadolinium-enhanced neutron tagging will 
first be done by SK.  {A corresponding project, SK-Gd, will start soon, after EGADS successfully 
demonstrated the sustainable gadolinium loading of water~\cite{Sekiya2016:SkGd, Watanabe2009:NeutTagWCGd}. 
}


The demonstrated feasibility to load common neutrino target materials with gadolinium is based on two 
positive properties: the large capture cross-section for thermal neutrons, especially of 
${}^{157}$Gd, and the high Q-value, 7937 keV~\cite{Choi2007:EGAF} for 
${}^{157}$Gd$(n,\gamma)$, compared to the values listed in Table~\ref{tbl:CommonCatcherIsotopes}.
The reason for the large cross-section of the gadolinium isotope 
is an s-wave neutron capture resonance state in the thermal 
energy region with a resonance energy of 31.4 meV for $^{157}$Gd \cite{CapGam_HP}. A list 
with the thermal neutron capture cross-sections of all the gadolinium isotopes in natural 
gadolinium, which defines the composition of how gadolinium is commonly loaded to 
neutrino target materials, is given in Table~\ref{tbl:GdNaturalAbundance}.\footnote{The thermal 
neutron capture cross-section of gadolinium, especially of ${}^{155}$Gd and ${}^{157}$Gd, is still 
under discussion~\cite{Leinweber2006:GdXSec, Choi2014:GdXSec}. }

\begin{table}[t!b]
    \centering
    \caption{Relative abundances of gadolinium isotopes in natural 
             gadolinium~\cite{Rosman1998:NatGd} and their radiative thermal neutron capture 
             cross-sections~\cite{Mughabghab2006:NeutRes}.}
    \label{tbl:GdNaturalAbundance}
    \begin{tabular}{rcc}
        \hline
        Isotope & Abundance & Cross-section\\
                & ${}$ [\%] &  [b]\\
        \hline
        ${}^{152}$Gd & 0.200 & 735    \\
        ${}^{154}$Gd & 2.18  & 85   \\
        ${}^{155}$Gd & 14.80 & 60900  \\
        ${}^{156}$Gd & 20.47 & 1.8   \\
        ${}^{157}$Gd & 15.65 & 254000 \\
        ${}^{158}$Gd & 24.84 & 2.2   \\
        ${}^{160}$Gd & 21.86 & 1.4   \\  
        \hline
    \end{tabular}
\end{table}


The about 8 MeV excitation energy from the Gd$(n,\gamma)$ reaction is released in  several
 $\gamma$ rays. Due to the calorimetric measurement, liquid scintillator detectors simply 
need to look for this energy deposition, assuming that all the $\gamma$ rays are fully contained 
inside the active 
volume. A water Cherenkov detector, however, detects only a part of it due to the 
above-mentioned energy threshold. 
Therefore, good 
understanding of the multiplicities of $\gamma$ rays from Gd$(n,\gamma)$ reactions and their energy 
distributions in the range 0.1-8 MeV is an important prerequisite to properly predict neutron 
tagging efficiencies in gadolinium-doped water Cherenkov detectors based on MC simulations. 

\section{Experiment}
\label{sec:Exp}

We performed our measurements of the thermal neutron capture on gadolinium with an enriched 
${}^{157}$Gd target inside the Ge spectrometer of 
ANNRI~\cite{Igashira2009:MLF-BL04, Kin2011:ANNRI, Kino2011:ANNRI, Kimura2012:ANNRI, Kino2014:ANNRI} 
at JSNS of J-PARC in December 2014. The JSNS complex provides neutrons with energies 
up to 100 keV. Its beam is one of the most intense pulsed neutron beams for precise neutron 
TOF experiments in the world, especially in the thermal energy region. The ANNRI 
detector, located at Beam Line No. 4~\cite{Igashira2009:MLF-BL04} of the MLF, is dedicated to 
measure cross-sections and $\gamma$-ray spectra of neutron-nucleus interactions 
with excellent energy resolution compared to other 
$\gamma$-ray spectrometers. 


\subsection{Detector Setup}
\label{subsec:Exps:ExpSetup}

During our measurements, the JSNS was powered by a 300 kW beam of 3 GeV protons in ``double-bunch 
mode'' that hit a mercury target at a repetition rate of 25 Hz. 
This created a double of 100 ns wide neutron beam bunches with 600 ns spacing every 40 ms. At the 
target position inside the ANNRI spectrometer, which is located 21.5 m  from the neutron beam 
source, the neutron beam delivered an energy-integrated neutron intensity of about $1.5\times10^7 / \mathrm{cm}^2 / \mathrm{s}$.

\begin{figure}[b!t]
   \begin{center}
      \includegraphics[width=12cm]{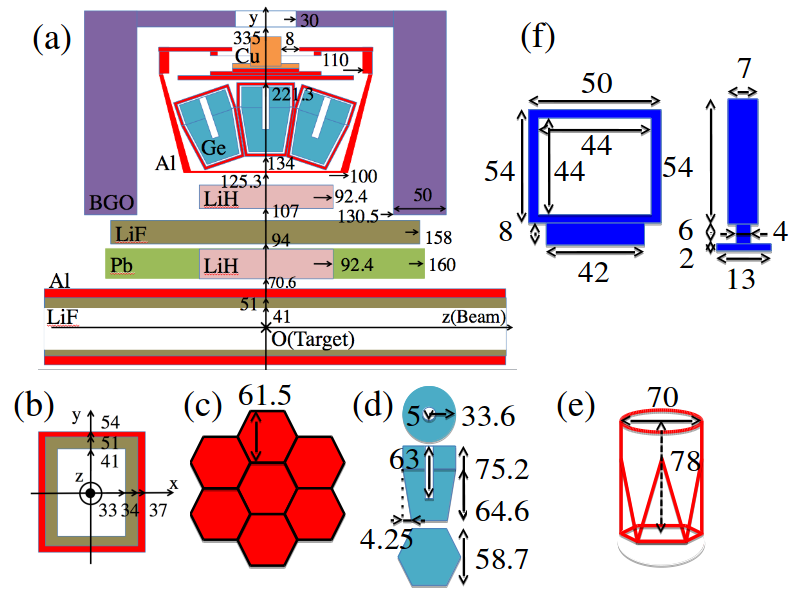}
      \caption{Schematics of parts of the ANNRI Ge spectrometer. Dimensions are given in 
       millimeters. (a) Upper cluster of Ge crystals (light blue) covered by an
       0.7 mm thick aluminum skin. It is 
      surrounded by BGO-Compton-suppression shields (purple). Further materials are copper 
      (orange), lead 
      (green), lithium fluoride (brown) and lithium hydride (pink). (b) Beam pipe profile. (c) 
      Hexagonal Ge cluster consisting of seven Ge crystals with hexagonal front faces 
      as seen from the target's perspective. (d) Breakdown of one Ge crystal. (e) Dimensions of  
      one Ge crystal. (f) Dimensions of the target holder.}
      \label{fig:detec}
   \end{center}
\end{figure}

The ANNRI spectrometer consists of two Ge cluster detectors with anti-coincidence shields 
made of bismuth Ge oxide (BGO) and eight co-axial Ge detectors. Since the co-axial 
detectors were still in repair after the Tohoku earthquake on 11 March 2011, we only used the two 
Ge cluster detectors shown in Fig.~\ref{fig:detec}(a) in the present analysis. The clusters are placed perpendicular 
to 
the aluminum beam pipe (Fig.~\ref{fig:detec}(b)), with the front faces 13.4 cm above and 
below the target position. They provide a combined solid angle coverage of 22\% 
with respect to this point. As shown in Fig.~\ref{fig:detec}(c) each of the 7 crystals in the cluster 
has its hexagonal surface facing the target.   
The dimensions of a Ge crystal are shown in Fig.~\ref{fig:detec}(d) 
and (e). 

The BGO anti-coincidence shield for one Ge cluster (see Fig.~\ref{fig:detec}(a)) consists 
of a cylindrical BGO counter, which is separated into twenty readout blocks: twelve 
around a cluster and eight covering its rear side. The shields provide a 
total solid angle coverage of 55\% with respect to the target.

In order to reduce background $\gamma$ rays from the neutron capture by the aluminium layer on the beam pipe, 
the inner face of the pipe is lined with a layer of lithium fluoride of $\sim$1~cm thickness. 
Moreover, shields made of lithium fluoride and lithium 
hydride are located between the pipe and the Ge clusters to protect the crystals from the 
impinging neutrons. The remaining $\gamma$-ray background was measured directly by placing only the 
empty target holder, whose dimensions are shown in Fig.~\ref{fig:detec}(f), inside the neutron beam.

\subsection {Data Acquisition}
\label{subsec:Exps:DAQ}

The data acquisition (DAQ) system~\cite{Kimura2008:ANNRIDAQ} was triggered when at least one of the 
fourteen Ge crystals had a collected charge equivalent of more than 100 keV. All 
further energy depositions in the crystals within a time window of 560 ns (smaller than the 
double-bunch spacing) after the trigger were combined with the initial deposition to form an event. 
Within this event, we only considered 
crystals with a collected charge corresponding to more than 100 keV as hit. The crystal hits 
of cluster were accepted if none of the 20 surrounding BGO blocks had an energy deposition greater 
than 100 keV within the same time window. The data stored per event included the neutron TOF, 
given by the time difference between the first detected hit of a crystal 
(trigger time) and a signal from the JSNS, as well as the collected charge (energy deposition) and 
the hit time delay with respect to the trigger time of every hit crystal.

For the purpose of dead time correction, signals from a random pulse generator with an average rate 
of 570 Hz were fed into the pre-amplifier of every Ge crystal and simultaneously  
counted by a fast counter. The amplitudes were set to be about an energy equivalent to 
9.5 MeV. The ratio $r_{L,i}$ (=$N_{r,i} / N_s$) of the number of pulses $N_{r,i}$ recorded by 
the $i$-th crystal to the number of  pulses $N_s$ corrects 
 the absolute elapsed time of the experiment $T$ 
for the dead time of the crystal's DAQ system after a trigger, giving the 
crystal's effective live time as $T_{L,i} = T \cdot r_{L,i}$. On average, $r_{L,i}$ is about 94\%. The 
dead time correction is important for calibration and background subtraction.

\subsection {Event classification}
\label{subsec:Exps:EventClass}

\begin{figure}[b!t]
  \centering
  \begin{minipage}{0.26\textwidth}
  \includegraphics[width=0.95\textwidth]{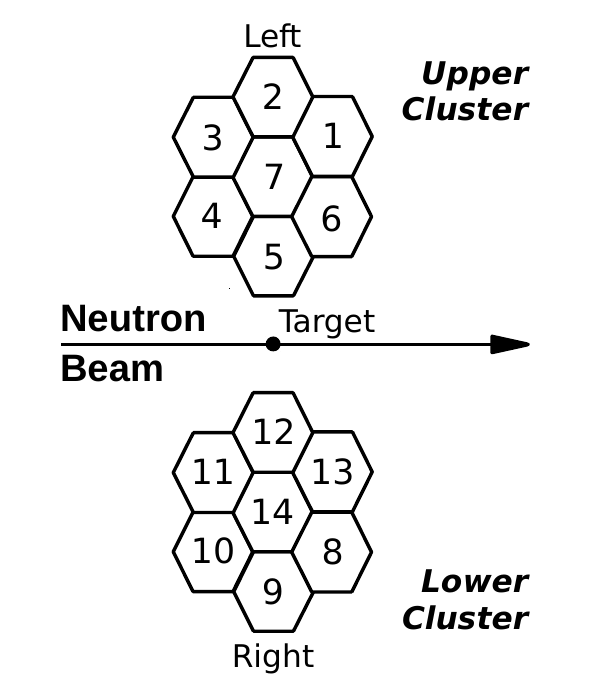}
  \end{minipage}%
  \hskip 3em
  \begin{minipage}{0.50\textwidth}
  \centering
  \includegraphics[width=0.90\textwidth]{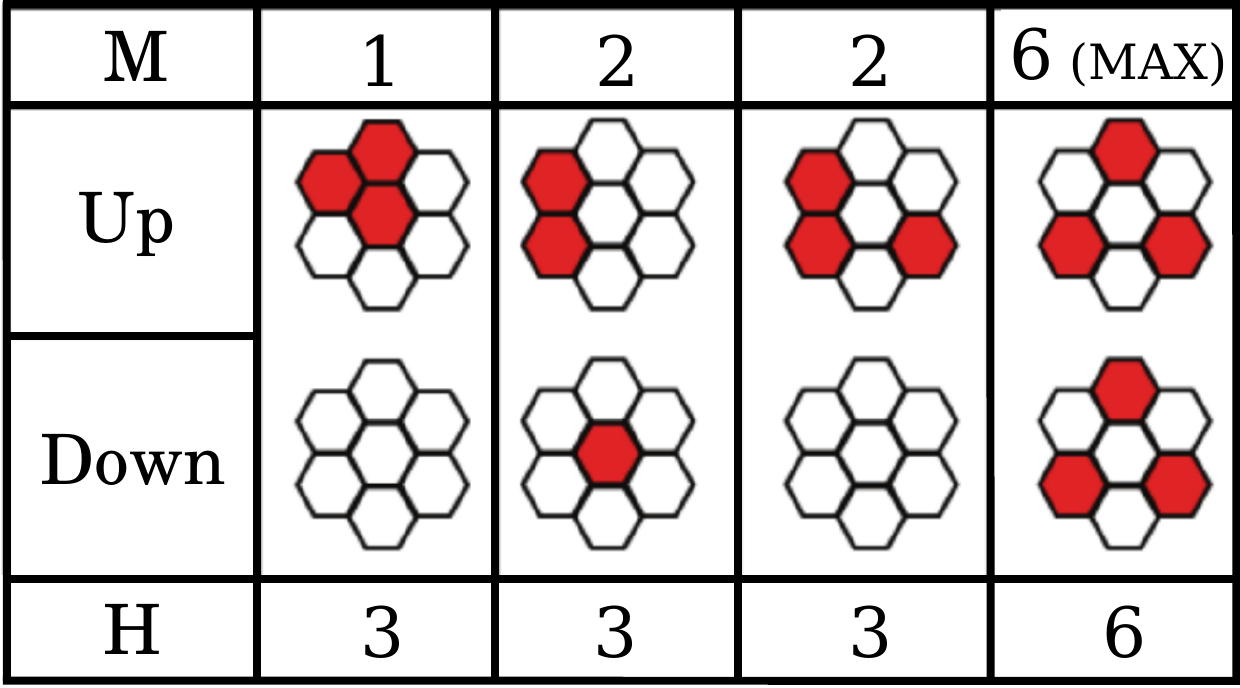}
  \end{minipage}%
  \caption{\textbf{Left}: Numbering scheme for the 14 Ge crystals of the upper and 
           the lower cluster. A cluster side being left (right) of the neutron beam in downstream 
           view is indicated by the annotation `Left' ('Right').
           \textbf{Right}: Examples of event crassification on how the multiplicity value M and the hit value 
           H of an event are assigned. From left to right one can see the following events: 
           M1H3 (means $\mathrm{M}=1$, $\mathrm{H}=3$), M2H3, M2H3 and M6H6.}
  \label{fig:multiplicity}
\end{figure}

We assigned a multiplicity value M and a hit value H to each recorded event. We defined the multiplicity 
M  as the combined number of isolated sub-clusters of hit Ge crystals at the upper 
and the lower  clusters. A sub-cluster is formed by the neighboring hit Ge crystals 
and can be of size $\geq 1$. The hit value H describes the total number of Ge crystals hit 
in the event. The multiplicity M represents the number of $\gamma$ rays and the hit value H represents 
the lateral spread of  $\gamma$ rays. 
 Figure~\ref{fig:multiplicity} shows some examples (right) together with the numbering 
scheme used to reference individual Ge crystals (left). 

Since we assume that M is the number of detected $\gamma$ rays, this implies that sub-clusters 
with sizes greater than one are mainly due to scattering of one $\gamma$-ray between neighboring crystals 
and not due to multiple $\gamma$ rays. 

\subsection{Detector simulation}
\label{subsec:Exps:DetSim}

Based on the geometry and material specifications for ANNRI (see Fig.~\ref{fig:detec}), 
we have developed a detailed detector simulation using version 9.6 patch 04 of the Geant4 toolkit. 
It uses the Geant4 implementation \textit{G4EmPenelopePhysics} of physics models for 
low-energy photon, electron-positron interactions developed for the PENELOPE (PENetration and 
Energy LOss of Positrons and Electrons) code version 2001~\cite{Geant4Physics2012:RefMan}.

With the MC simulation we evaluated the detector response to the simultaneous propagation of one 
or more $\gamma$ rays with specified energies through the setup. During the simulation of an
event, each Ge crystal accumulated the energy depositions from charged particles. This 
information was then used to realize the trigger and veto scheme described in 
Sect.~\ref{subsec:Exps:DAQ}. We validated the MC simulation by comparing its outcomes for the 
fractions of different event 
classes (Sect.~\ref{subsec:Exps:EventClass}) and the energy spectra observed by the single 
crystals to the data taken with calibration sources (Sect.~\ref{subsec:Exps:CalibData}): 
Radioactive $^{60}$Co dominantly emits two $\gamma$ rays, 1173 keV and 1332 keV, 
after its $\beta^-$ decay to $^{60}$Ni. We used the lower cluster of ANNRI to tag one of the 
$\gamma$ rays in a single crystal and looked at the crystal hit configuration created by the other 
$\gamma$ ray in the upper cluster. This resulting hit configuration was classified with multiplicities $\mathrm{M}=1,2$ and hit 
values $\mathrm{H}=1,2,3$. The tagging of one $\gamma$ ray with the lower cluster ensures that the 
upper hit configuration stems solely from the other $\gamma$ ray. 

To study the hit configurations at higher energy, we used the single
$\gamma$ ray  of 8579 keV from the thermal neutron capture ${}^{35}$Cl$(n,\gamma)$ reaction, which is produced 
via direct M1/E2 transition from 8579 keV  to the ground state ($2^{+} \rightarrow 2^{+}$) \cite{Mughabghab2006:NeutRes}. Table~\ref{tab:Hits} summarizes the fractions of the different event classes created by the 
$\gamma$ rays of different energies in our experimental data and our MC simulation. 
We only selected events with $\mathrm{M}=1$, i.e., with one 
sub-cluster of hit crystals. 
Using the MC simulation, we estimated the background contribution which comes mainly from 6 prominent two-step deexcitation
$\gamma$ rays from 8579 keV using the MC simulation~\cite{Cl35:Capture} to be about 9\% for M1H2 case and 24\% for 
M1H3 case. The table lists the values after subtracting these contributions. The systematic errors to the numbers given in the table 
due to this overlap effect are negligible. 


As one can see from Table~\ref{tab:Hits},  
the agreement between data and MC for the two $^{60}$Co lines is very good. Despite the  
errors for the experimental data on the 8579 keV line from ${}^{36}$Cl due to the above 
corrections, the agreement with the MC simulation is also good. 
Overall, the summary shows that energy migration to neighboring and distant crystals within a 
Ge cluster, which increases with increasing $\gamma$-ray energy and arises from 
Compton scattering of the $\gamma$ ray or  $\gamma$-induced 
$e^+e^-$ pair production, is correctly reproduced within our MC simulation.  

\begin{figure}[b!t]
  \centering
  \begin{minipage}{0.49\textwidth}
  \includegraphics[trim=0.1cm 0.1cm 1.0cm 0.1cm, clip=true, 
  width=\textwidth]{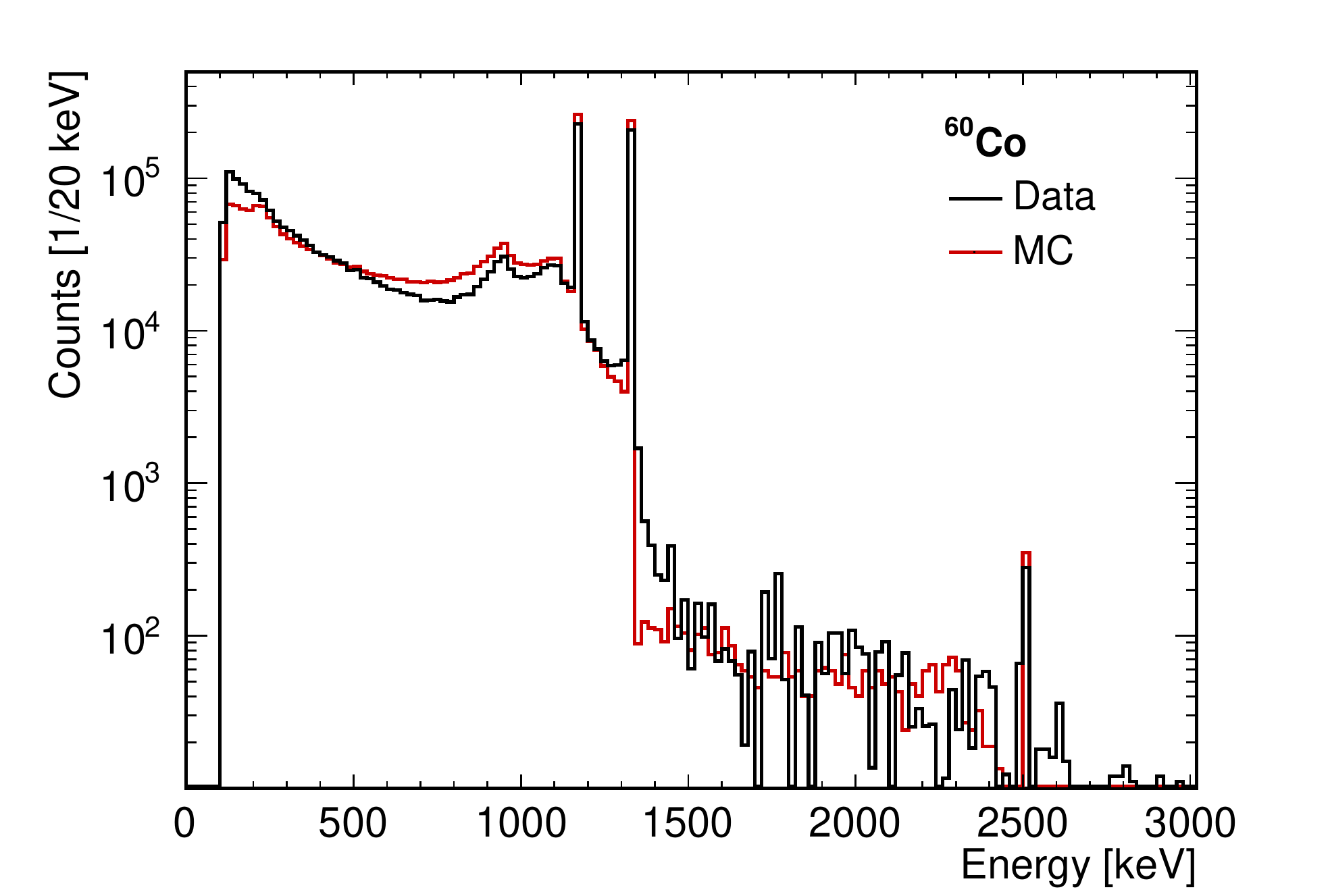}
  \end{minipage}%
  ~
  \begin{minipage}{0.49\textwidth}
  \includegraphics[trim=0.1cm 0.1cm 1.0cm 0.1cm, clip=true, 
  width=\textwidth]{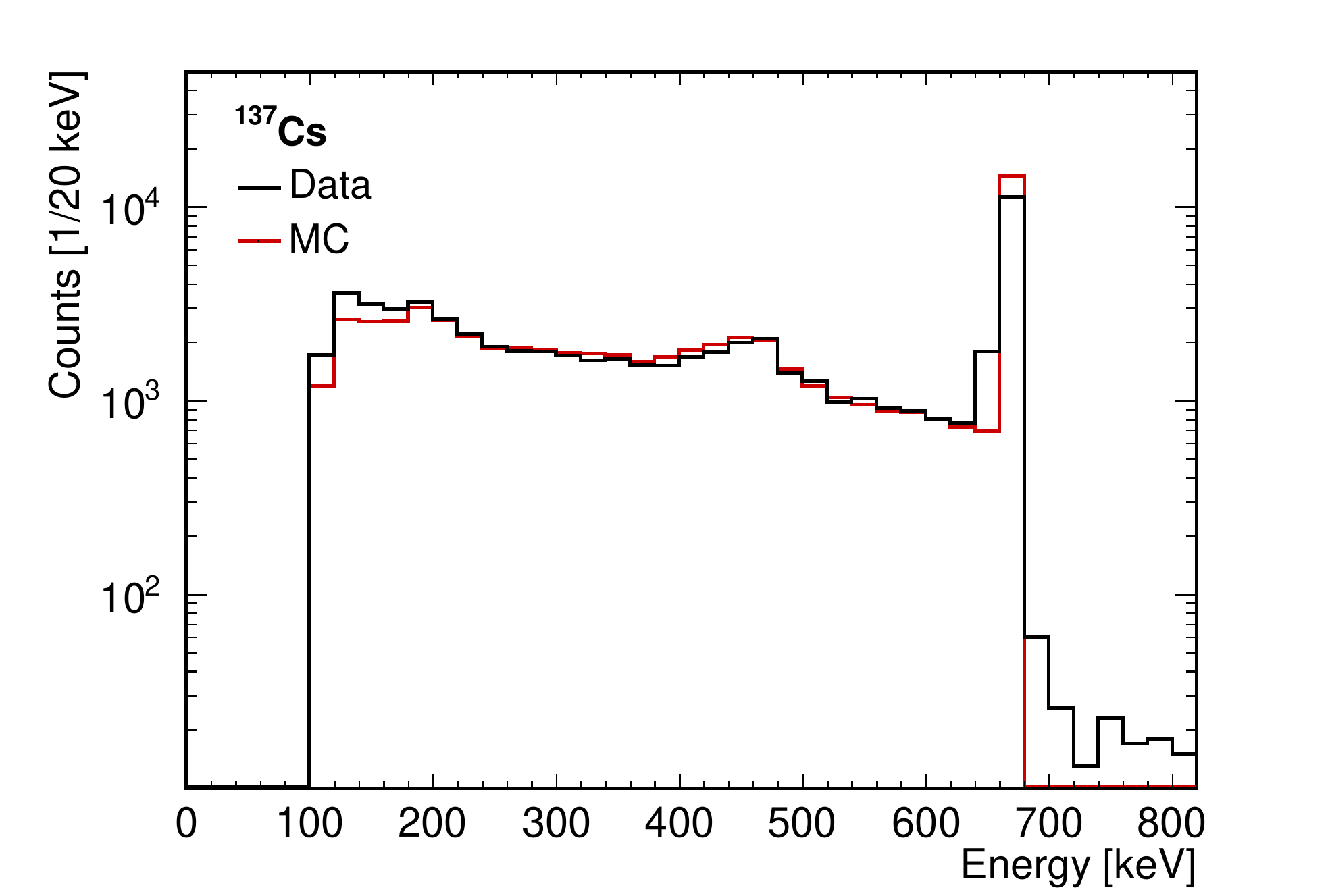}
  \end{minipage}%
  \caption{Energy spectra from M1H1 events observed by the peripheral crystal 6 of the upper cluster
           in our data (black) and our MC (red) for the calibration sources $^{60}$Co (left) and 
           $^{137}$Cs (right).}
  \label{fig:CalibDataVsMc}
\end{figure}

Moreover, Fig.~\ref{fig:CalibDataVsMc} shows the energy spectra for $^{60}$Co (left) and $^{137}$Cs 
(right) from M1H1 events observed in our calibration data and corresponding MC. One can see that, 
in addition to the multiplicities, also the spectral shapes are very well reproduced by our 
detector simulation.


\begin{table}[t!b]
    \centering
    \caption{Summary of the fractions of different event classes ($\mathrm{M}=1,2$, 
             $\mathrm{H}=1,2,3$) created by three different, single $\gamma$ rays in our 
             experimental data (Exp) obtained with calibration sources / targets ($^{60}$Co and 
             ${}^{35}$Cl) and in our MC simulation. 
             Errors are from statistics only.}
    \label{tab:Hits}
    \begin{tabular}{c|c||c||ccc}
    
      \hline 
      \multicolumn{2}{c||}{Class} & Data & \multicolumn{3}{c}{Fraction [\%]}\\
      \hline
      M & H &  & 1173 keV & 1332 keV & 8579 keV\\
      \hline \hline
      \multirow{6}{*}{1} & \multirow{2}{*}{1} & Exp & $71.2\pm0.3$ & $70.7\pm0.3$ & $48.6\pm1.4$\\
                         &                    & MC  & $71.5\pm1.1$ & $70.0\pm1.1$ & $46.5\pm0.3$\\
      \cline{2-6}
                         & \multirow{2}{*}{2} & Exp & $26.3\pm0.3$ & $26.7\pm0.3$ & $37.7\pm1.9$\\
                         &                    & MC  & $26.1\pm0.6$ & $27.4\pm0.6$ & $39.5\pm0.3$\\
      \cline{2-6}
                         & \multirow{2}{*}{3} & Exp & $2.00\pm0.09$& $2.2\pm0.1$  & $11.6\pm2.1$\\
                         &                    & MC  &  $2.1\pm0.1$ & $2.2\pm0.1$  & $11.6\pm0.1$\\
      \hline
      \multirow{4}{*}{2} & \multirow{2}{*}{2} & Exp & $0.40\pm0.04$& $0.40\pm0.04$& --\\
                         &                    & MC  & $0.38\pm0.06$& $0.40\pm0.06$& --\\
      \cline{2-6}
                         & \multirow{2}{*}{3} & Exp & $0.03\pm0.01$& $0.010\pm0.007$& --\\
                         &                    & MC  & $0.02\pm0.01$& $0.02\pm 0.01$&--\\
      \hline   
    \end{tabular}
\end{table}


\subsection{Background and calibration data}
\label{subsec:Exps:CalibData}

\begin{figure}[b!t]
  \begin{center}
    \includegraphics[trim=0.1cm 0.1cm 1.0cm 0.1cm, clip=true,width=0.55\textwidth] 
    {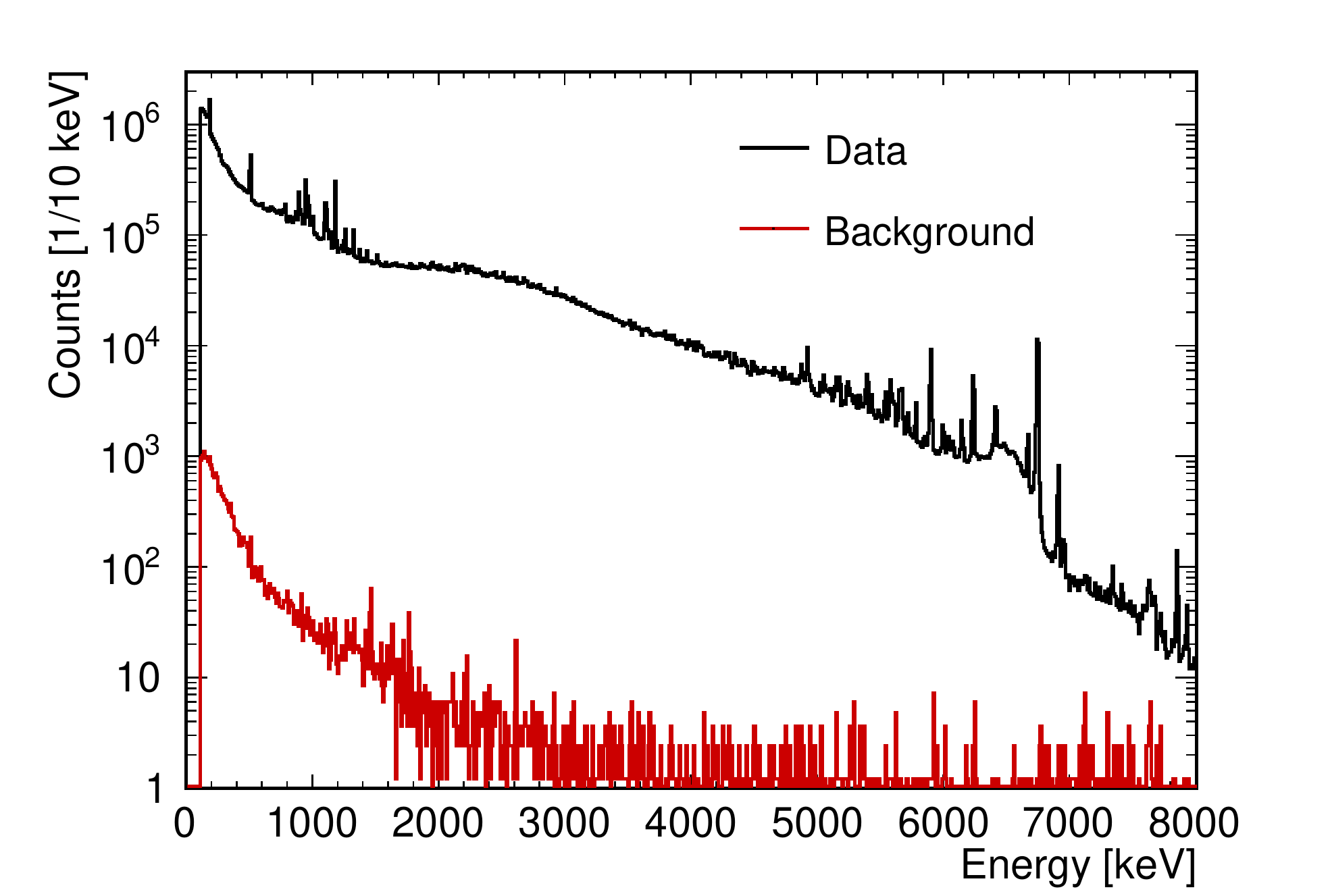}
    \caption{Energy spectra for M1H1 events observed by crystal 6 in the background measurement 
             with an empty target holder (red) and the measurement with the enriched ${}^{157}$Gd 
             sample (black; before background subtraction). The 
             background spectrum was scaled to match the dead-time-corrected live time of the 
             gadolinium measurement.
             }
    \label{fig:BkgSpec}
  \end{center}
\end{figure}

In order to measure the background for the experiment, which originates mostly from 
$\gamma$ rays from the interactions of the beam neutrons with materials other than the 
 target, we placed the empty target holder for 6 hours into the neutron beam. 
Figure~\ref{fig:BkgSpec} shows 
the background energy spectrum observed by one of the crystals for M1H1 events together with the spectrum 
observed in the measurement with the enriched ${}^{157}$Gd sample before background subtraction. The background spectrum, 
after processing in the same way as the data and the live-time normalization, 
contributes only  $\sim$0.06\% to the gadolinium data spectrum.

The energy calibration of the ANNRI Ge crystals was done with known $\gamma$-ray lines 
from the radioactive sources $^{60}$Co, $^{137}$Cs and $^{152}$Eu as well as from the deexcitation 
of $^{36}$Cl after the thermal $^{35}$Cl($n$,$\gamma$) reaction in a sodium chloride (NaCl) target. 
Table~\ref{tab:EvtCounts} summarizes the measurement time and number of observed events for the 
different sources and targets.

\begin{table}[b!t]
   \centering
   \caption{Data sets recorded for calibration with different sources (left; without beam) and 
            neutron beam targets (right). The line `Empty' means that only the empty target 
            holder was placed inside the neutron beam for a background measurement.}
   \label{tab:EvtCounts}
   \begin{tabular}{r|cc||r|cc}
      \hline
      Source     & Time        & Events            & Target & Time      & Events \\
      \hline 
      &&&&&\\[\dimexpr-\normalbaselineskip+0.1em]
      $^{60}$Co  & 18 h  & $8.8 \times 10^7$ & NaCl   & 4 h & $1.3 \times 10^8$\\[0.1em]
      $^{137}$Cs & 0.5 h & $2.1 \times 10^6$ & Empty  & 6 h & $1.3 \times 10^7$\\[0.1em]
      $^{152}$Eu & 7 h   & $2.3 \times 10^7$ &        &           &\\
      \hline
   \end{tabular}
\end{table}


The energy resolutions ($\sigma (E)$) of all the 14 crystals were measured over the energy from 0.3 to 8 MeV and 
they are expressed as $\sigma (E) ({\rm keV})=1.8+0.00041E ({\rm keV})$. 


With the known activities $\beta$ of our $^{60}$Co, $^{137}$Cs and $^{152}$Eu sources,
we estimated absolute single photopeak efficiencies $\varepsilon_{i}(E_\gamma)$ at different energies
$E_\gamma$ for each crystal ($i$) as
\begin{equation}
 \varepsilon_{i}(E_\gamma) = \frac{N_i(E_\gamma)}{\mathrm{BR}_\gamma \, \beta T_{L,i}} \, ,
\end{equation}
where  $N_i(E_\gamma)$ is the number of detected $\gamma$ rays in the $\pm3\sigma$ region of a Gaussian fitted 
to the photopeak observed by the $i$-th crystal at energy $E_\gamma$, 
$\mathrm{BR}_\gamma$ is the branching ratio for the decay branch emitting the $\gamma$ ray 
of energy $E_\gamma$, and $T_{L,i}$ is the corrected livetime. 

\begin{figure}[b!t]
  \begin{center}
    \includegraphics[trim=0.1cm 0.1cm 1.0cm 0.1cm, clip=true,width=0.55\textwidth] 
    {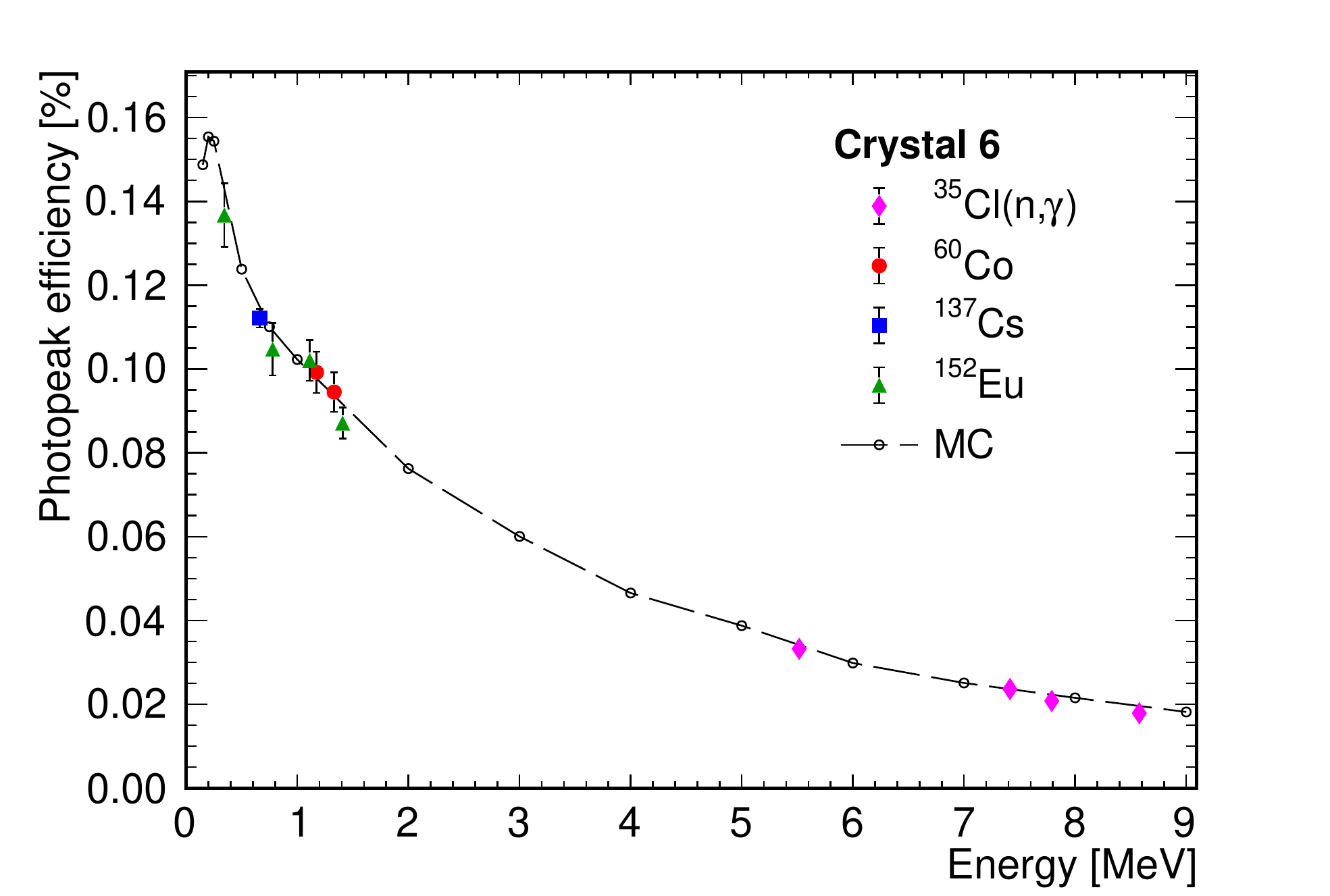}
    \caption{Single photopeak efficiencies at different $\gamma$-ray energies for the peripheral 
          crystal 6 of the upper Ge  cluster in ANNRI. 
          The data points are from measurements with the radioactive $^{60}$Co (1173 keV, 
          1332 keV), $^{137}$Cs (662 keV) and $^{152}$Eu (344 keV, 779 keV,
          1112 keV, 1408 keV) sources and of the thermal 
          ${}^{35}$Cl$(n,\gamma){}^{36}$Cl reaction (5517 keV, 7414 keV, 
          7790 keV, 8579 keV) with the NaCl target. Points named `MC' are the single photopeak MC efficiencies from Eq.~(\ref{eq:MCeff}).}
    \label{fig:PhotopeakEff}
  \end{center}
\end{figure}

The single photopeak efficiency values at various energies from the measurements 
with the radioactive sources and the NaCl target cover the range from 344 keV to 
8579 keV for each crystal. The values for one of the crystals are depicted in Fig.~\ref{fig:PhotopeakEff}. The relative efficiency values for 
the NaCl target were normalized with respect to the dominant 7414 keV line, which 
itself was normalized with our MC simulation. 


The corresponding prediction for each crystal  ($i$) was calculated using the MC simulation as 
\begin{equation}
\varepsilon^{\mathrm{MC}}_{i}(E_\gamma) = \frac{N_i(E_\gamma)}{N(E_\gamma)} \, ,
\label{eq:MCeff}
\end{equation}
where $N_i(E_\gamma)$ is the number of $\gamma$ rays and $N(E_\gamma)$ 
denotes the total number of generated $\gamma$ rays with energy $E_\gamma$. 
The data points and the MC simulation are in good agreement. 

\begin{figure}[b!t]
  \begin{center}
    \includegraphics[trim=0.1cm 0.1cm 1.0cm 0.1cm, clip=true,width=0.55\textwidth] 
    {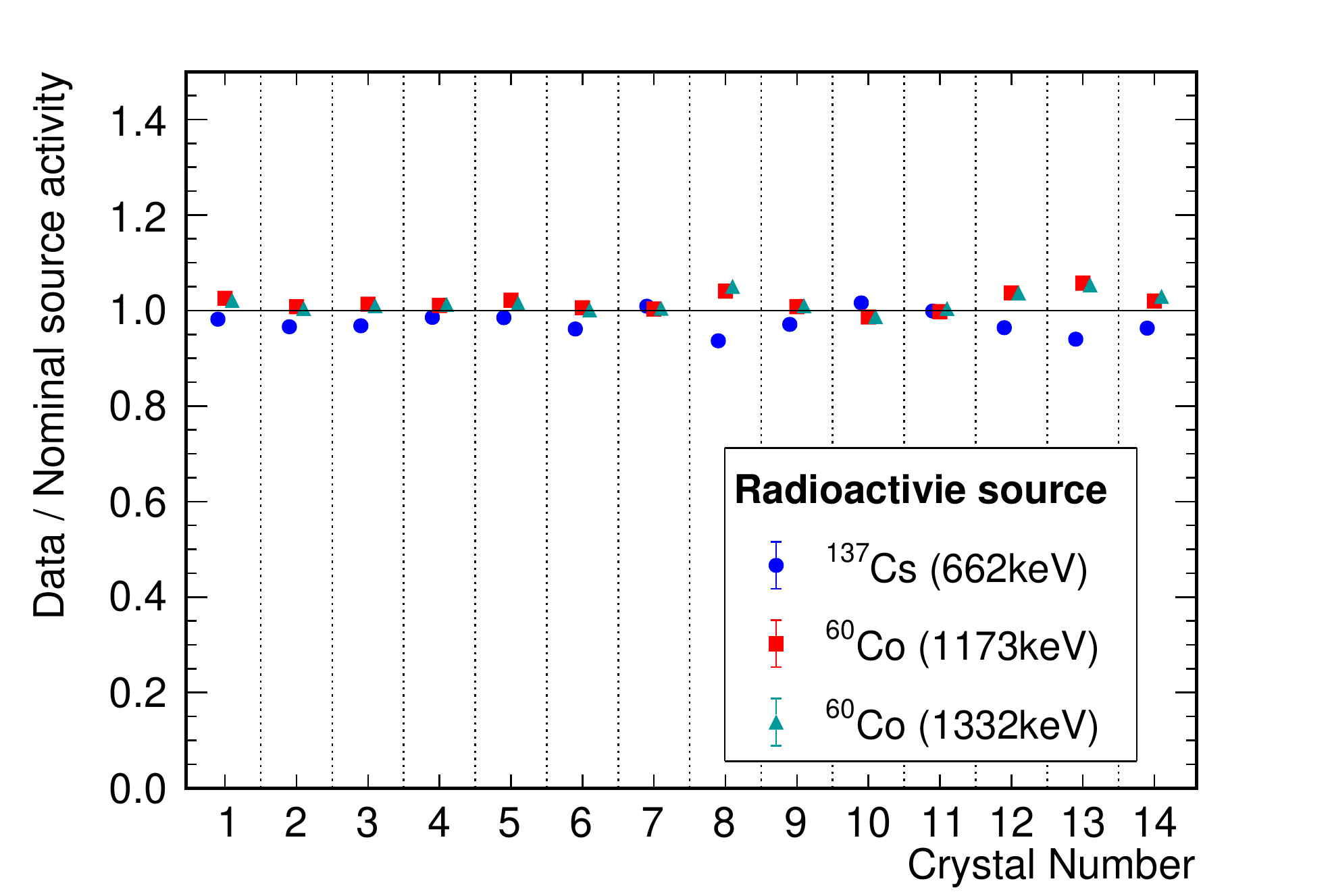}
    \caption{Ratios of the measured activities for the ${}^{60}$Co and $^{137}$Cs 
             calibration sources to the sources' corresponding nominal values 
            ($\beta_{\mathrm{Co}}=4850$ Bq, $\beta_{\mathrm{Cs}}=6317$ Bq) for each 
            Ge crystal in the ANNRI detector. 
            }
    \label{fig:ActivityRatios}
  \end{center}
\end{figure}

The data from the $^{60}$Co and $^{137}$Cs calibration sources also allowed us to check the 
uniformity of the detector. For this purpose, we compared the nominal value of the source's activity to the 
value measured by each Ge crystal. The ratios of data to nominal value are 
shown in Fig.~\ref{fig:ActivityRatios}. 


Taking the error bars into account, the spread of the single ${}^{60}$Co ($^{137}$Cs) ratios with 
respect to the mean ratio shows a uniformity of the detector response over the solid angle of the detector 
at the 8\% (14\%) level. In other words, the detection efficiency is well understood over all crystals at this level of uniformity. 
Further details are described in Appendix~\ref{sec:AppEfficiency}.

\subsection {Gadolinium data}
\label{subsec:Exps:GdData}

For the measurement with gadolinium we attached the enriched ${}^{157}$Gd target in the form of 
gadolinium oxide powder (Gd$_2$O$_3$) in a teflon sheet to the designated holder within the neutron beam 
line at the center of the ANNRI detector. 
Taking into account the isotopic composition of the commercial gadolinium 
sample (Table~\ref{tab:GdOxPowder}) and the dominant cross-section 
of ${}^{157}$Gd (see Table~\ref{tbl:GdNaturalAbundance}), the target is essentially a pure 
${}^{157}$Gd target for thermal neutrons. A total of $1.81 \times 10^9$ events were collected with 
this target in about 44 hours of data taking.

\begin{table}[b!t]
	\centering
	\caption{Relative abundances of gadolinium isotopes in the Gd$_2$O$_3$ powder 
                 target~\cite{Isoflex2014:Certificate}.}
	\label{tab:GdOxPowder}
	\begin{tabular}{r|ccccccc}
		\hline
		Gd Isotope     &152      & 154  & 155  & 156  & 157            & 158  & 160\\
                \hline
                Abundance [\%] & $<0.01$ & 0.05 & 0.30 & 1.63 & $(88.4\pm0.2)$ & 9.02 & 0.60 \\
		\hline
	\end{tabular}
\end{table}

\begin{figure}[b!t]
  \begin{center}
    \includegraphics[trim=0.1cm 0.0cm 1.0cm 0.1cm, clip=true,width=0.55\textwidth] 
    {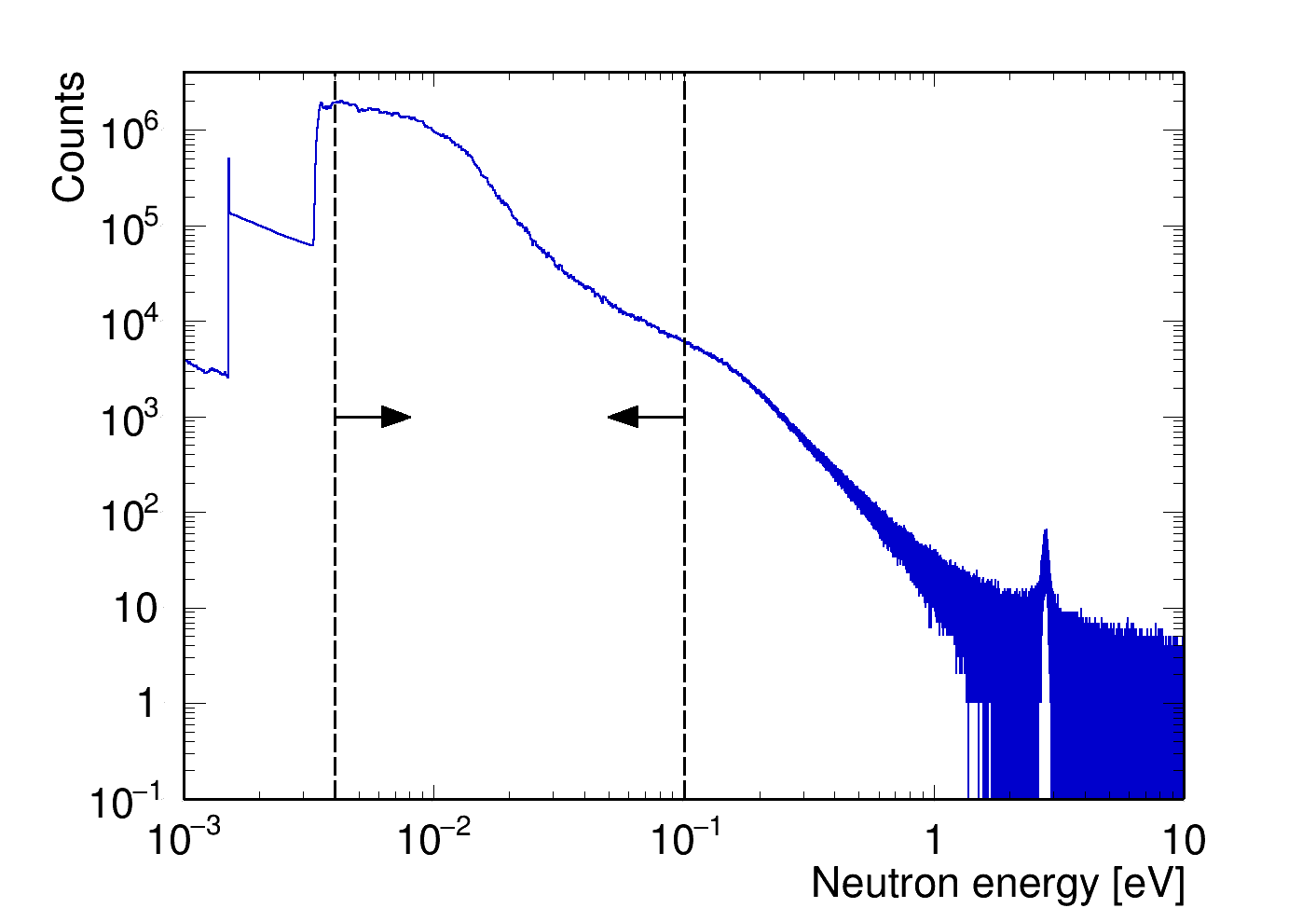}
    \caption{Energy spectrum of neutrons as obtained with the observed 
             neutron TOF according to Eq.~(\ref{eq:NeutronToFtoErgy}).}
    \label{fig:NSpec}
  \end{center}
\end{figure}

From the neutron TOF $T_{\mathrm{TOF}}$ recorded for each event 
we calculated the neutron kinetic energy $E_n$ as
\begin{equation}
  E_n = m_n(L/T_\mathrm{TOF})^2 / 2 \, ,
  \label{eq:NeutronToFtoErgy}
\end{equation}
where $m_n$ is the neutron mass and $L$ is the 21.5 m distance 
between neutron source and target. The resulting neutron energy spectrum is shown in 
Fig.~\ref{fig:NSpec}. 
In order to study the $\gamma$-ray 
spectrum solely from thermal neutron capture on ${}^{157}$Gd, we only selected events from neutrons 
in the kinetic energy range $[4, 
100]$ meV for the present analysis.

\begin{figure}[b!t]
  \centering
  \begin{minipage}{0.49\textwidth}
  \includegraphics[trim=0.1cm 0.1cm 1.0cm 0.1cm, clip=true, 
  width=\textwidth]{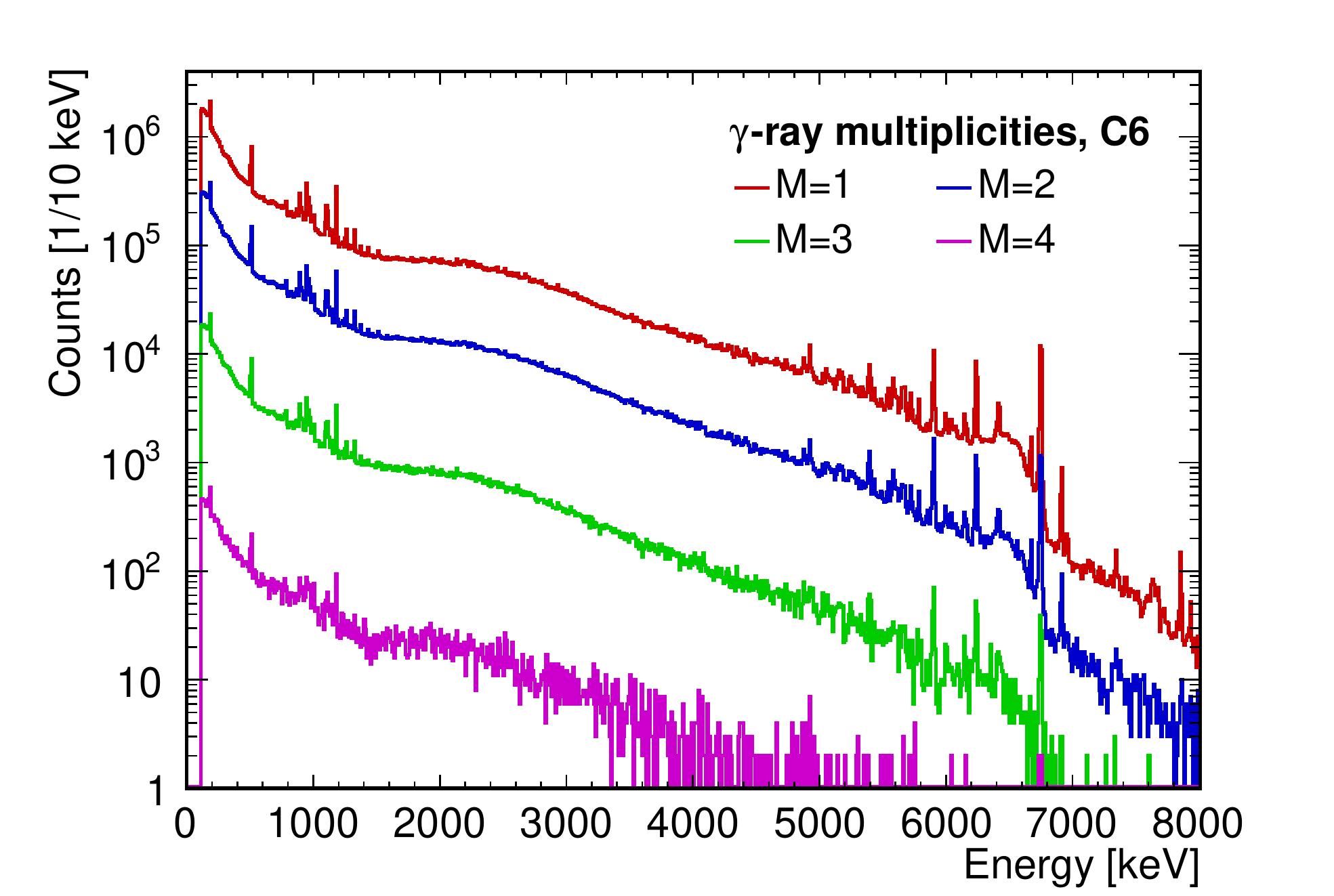}
  \end{minipage}%
  ~
   \begin{minipage}{0.49\textwidth}
  \includegraphics[trim=0.1cm 0.1cm 1.0cm 0.1cm, clip=true, 
  width=\textwidth]{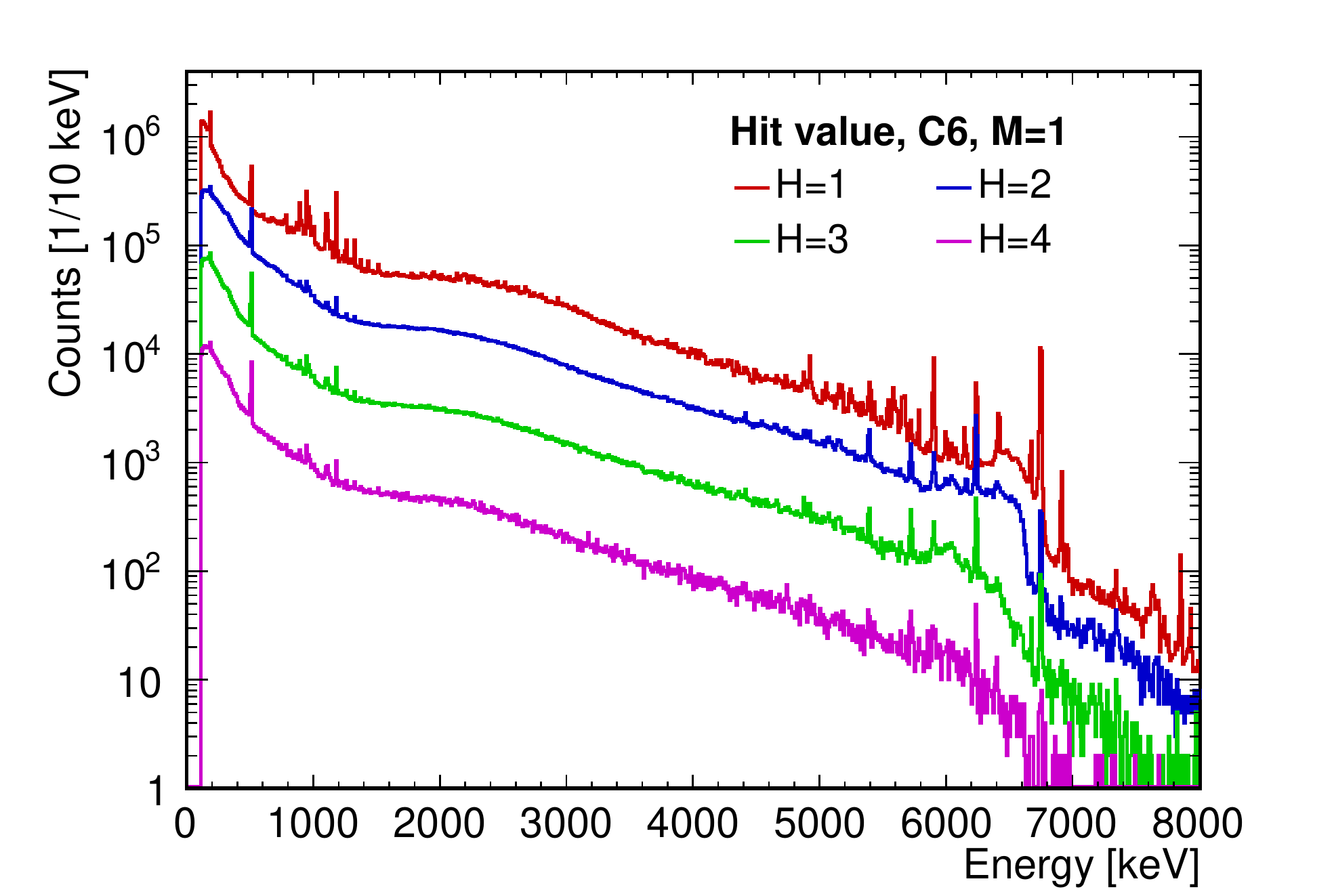}
  \end{minipage}%
  \caption{Energy spectra from thermal ${}^{157}$Gd$(n,\gamma)$ events 
  with: (i) Different, assigned $\gamma$-ray 
          multiplicities M 
          and (ii) Different hit values H with assigned $\gamma$-ray multiplicity 
          $\mathrm{M}=1$ (right)
          that were observed by the peripheral crystal 6 
          of the upper  cluster in ANNRI. From top to bottom the assigned multiplicities are 
          $\mathrm{M}=1$, 2, 3 
          and 4 (left) and assigned hit values are 
          $\mathrm{H}=1$, 2, 3 and 4 (right).}
  \label{fig:multispectra}
\end{figure}

After the neutron energy selection and the subtraction of the background,  
the resulting 
event sample was divided into sub-samples based on the multiplicity M and hit value H of the 
events. Figure~\ref{fig:multispectra} shows the energy spectra observed by the 
crystal 6  for different multiplicity values M (M1H1, M2H2 and M3H3). 
We mainly show the spectra from the hit configurations  M1H1, M2H2 and M3H3, 
since they are the majority of the events among each multiplicity value (M=1, 2 and 3)  and they are 
less subject to the overlap with multiple $\gamma$ rays.


The observed spectra 
are consistent within about 7\% for the dominant M1H1 
events for all 14 detectors. 
The observed energy spectra are dominated by the $\gamma$ rays from the thermal 
${}^{157}$Gd$(n,\gamma)$ reaction,  especially when we selected M1H1 and M2H2 events, since a clean single hit
on one crystal suppresses the effect of Compton scattering. 
At low energy, the spectra are slightly distorted by the effect of  the Compton scattering.  


The M1H1 spectra in Fig.~\ref{fig:multispectra} exhibit two components: discrete peaks, very well 
visible below 1.6 MeV and above 4.8 MeV,\footnote{Note 
that the spectra in Fig.~\ref{fig:multispectra} contain single and double escape peaks in addition to 
the relevant photopeaks.} and a continuum, most prominent between the previous energy regions. The 
origins and features of these components and how we implemented them in our spectrum model will be 
discussed in the following sections.

\section{Gamma-rays from thermal ${}^{157}$Gd$(n,\gamma)$ reaction: Emission scheme and model}
\label{sec:Model}


Our approach to model the $\gamma$-ray spectrum is a separate description of the continuum 
component and the discrete peaks visible in Fig.~\ref{fig:multispectra}.
 We followed the strategy of the GLG4sim package~\cite{GLG4sim2005:HP} for 
Geant4 to which we compare our results in Sect.~\ref{sec:Performance}. 


\subsection {Emission scheme}
\label{subsec:Model:EmScheme}

After the thermal neutron capture on ${}^{157}$Gd, the remaining ${}^{158}$Gd${}^*$ compound 
nucleus is in an s-wave neutron capture resonance state with an excitation energy of 7937 keV and spin-parity $J^{\pi}= 2^-$ 
\cite{Mughabghab2006:NeutRes}. It deexcites via a cascade of on average four $\gamma$-ray 
emissions~\cite{Chyzh2011:DANCE} to 
the ground state of  ${}^{158}$Gd with $J^{\pi}={0}^+$. 

\begin{figure}[b!t]
  \centering
  \begin{minipage}{0.55\textwidth}
  \raggedleft
  \includegraphics[height=6cm]{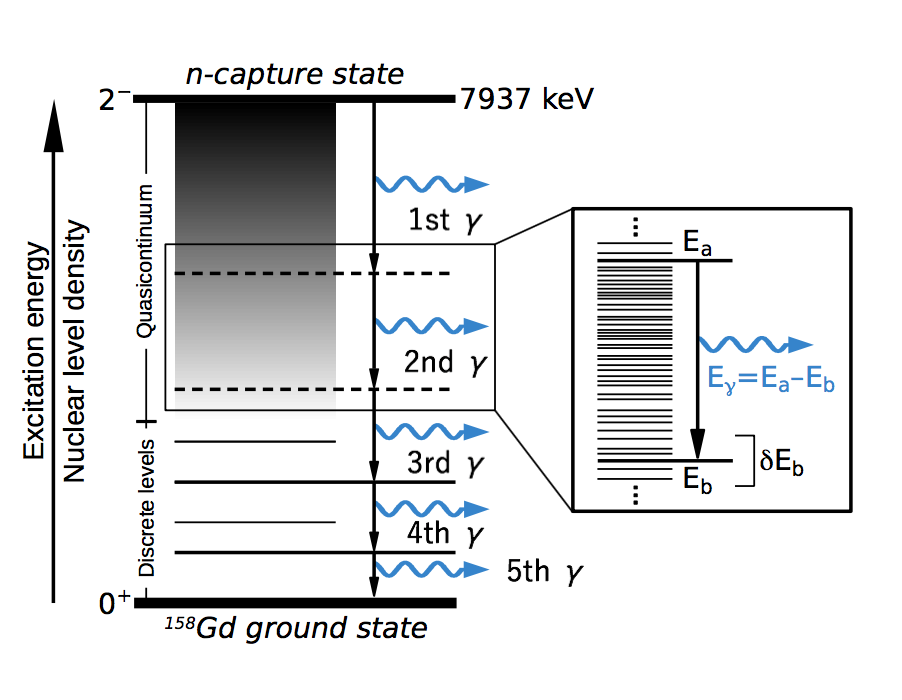}
  \end{minipage}%
  \hskip 1em
  \begin{minipage}{0.38\textwidth}
  \raggedright
  \includegraphics[height=6cm]{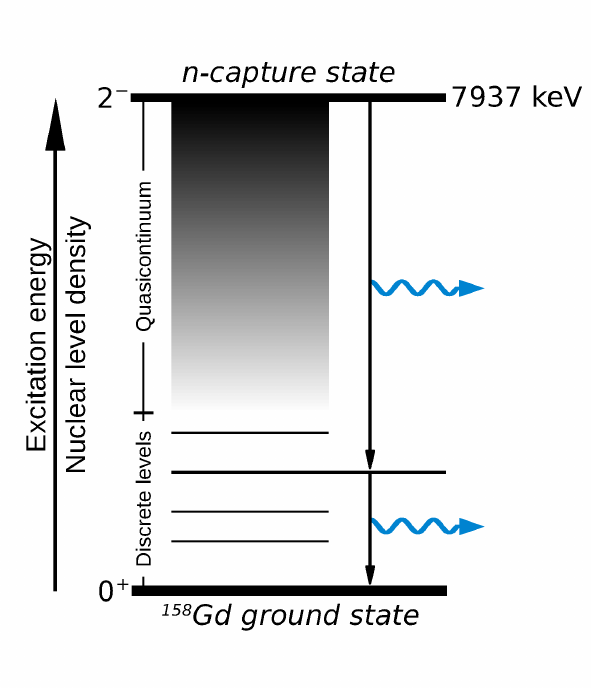}
  \end{minipage}%
  \caption{\textbf{Left}: Illustration of a multi-step $\gamma$-ray cascade from the neutron 
           capture (n-capture) state down towards the ground state via many intermediate 
           levels in the deexcitation of ${}^{158}$Gd${}^*$ after the thermal 
           ${}^{157}$Gd$(n,\gamma)$ reaction.
           \textbf{Right}: Example for a two-step cascade that proceeds via a low-lying level and
           includes the emission of a high-energy $\gamma$ ray. This cascade type contributes to 
           the creation of discrete peaks in the high-energy part of the $\gamma$-ray spectrum.}
  \label{fig:EmissionsModes}
\end{figure}

As illustrated on either side of 
Fig.~\ref{fig:EmissionsModes}, the density of nuclear levels increases with increasing excitation 
energy from the domain of well separated (discrete) levels, where spin and 
parity of the states are known, to a quasicontinuum where individual states and energy levels 
cannot be resolved. Since the two regions are connected smoothly, there is no obvious, sharp 
boundary between them. For the purpose of modeling, an arbitrary transition point is commonly 
defined at an excitation energy up to which supposedly complete information on discrete levels is 
available, e.g., 2.1 MeV in Ref.~\cite{Chyzh2011:DANCE}.

As depicted on the left of Fig.~\ref{fig:EmissionsModes}, the continuum component of the 
$\gamma$-ray spectrum from the thermal ${}^{157}$Gd$(n,\gamma)$ reaction stems from multi-step 
deexcitations of ${}^{158}$Gd${}^*$. 
Such 
intermediate transitions from the neutron capture state down towards the ground state can occur 
between (unresolvable) levels in the quasicontinuum (dashed lines), within the domain of 
discrete levels (solid lines) and between two levels from each of these smoothly connected regions. 
Both the number and energy values of the emitted $\gamma$ rays (i.e., the intermediate levels) are 
random. 

The discrete peaks on top of the continuum mainly originate from the transition from the neutron capture state 
to the low-lying levels  as illustrated on the right 
of Fig.~\ref{fig:EmissionsModes} and they are studied in the previous publications~\cite{Ali1994:Gd157, Bollinger1970:NeutARC}.
While the discrete peaks in our model are based on the previous publications and their intensities are 
adjusted to agree with our own data, we employ a statistical approach to describe the 
continuum component in the $\gamma$-ray 
energy spectrum of ${}^{158}$Gd${}^*$, which dominates with a contribution of $(93.06\pm 0.01)\%$ to 
our data. The approach is to follow Fermi's Golden 
Rule~\cite{Fermi1950:Book}, which states that the transition probability per 
unit time is proportional to the product of the transition matrix element squared between the initial and 
the final states and the state density at the final state:
Starting from an excited state with energy $E_a$, the differential probability 
$\mathrm{d}P(E_a,E_b)/\mathrm{d}E$ that the nucleus undergoes a transition to a state with energy 
$E_b (< E_a)$ and emits a $\gamma$ ray of 
multipolarity $XL$ ($X=E,M$ and $L=1,2,\dots$ for electric, magnetic and angular momentum, respectively) with energy $E_{\gamma}=E_a-E_b$ is expressed as 
\begin{equation}
  \frac{\mathrm{d}P(E_a,E_b)}{\mathrm{d}E} \propto \rho(E_b) \times \sum \limits_{XL} 
 \underbrace{2\pi \, E_{\gamma}^{2L+1} \, 
 f_{XL}(E_{\gamma})}_{T_{XL}(E_{\gamma})} \, .
\label{eq:continuum_prob_base}
\end{equation}
The first factor, $\rho(E_b)$, is the nuclear level density (NLD) at the final state ($E_b$)~\cite{Capote2009:RIPL3}. The 
second factor is the sum over the transmission 
coefficients $ T_{XL}(E_{\gamma})$ for the different multipolarities, each of them depending on 
the corresponding photon strength function (PSF), $f_{XL}(E_{\gamma})$, for 
electromagnetic decay. 

Since $\rho(E_b)$ increases exponentially as  $E_b$ increases, this factor favors transitions from  $E_a$ to  $E_b$ 
which is large and close to $E_a$, and thus favors  small $E_\gamma=E_a-E_b$. 

\begin{figure}[b!t]
  \centering
  \begin{minipage}{0.49\textwidth}
  \includegraphics[trim=0.1cm 0.1cm 1.0cm 0.1cm, clip=true, 
  width=\textwidth]{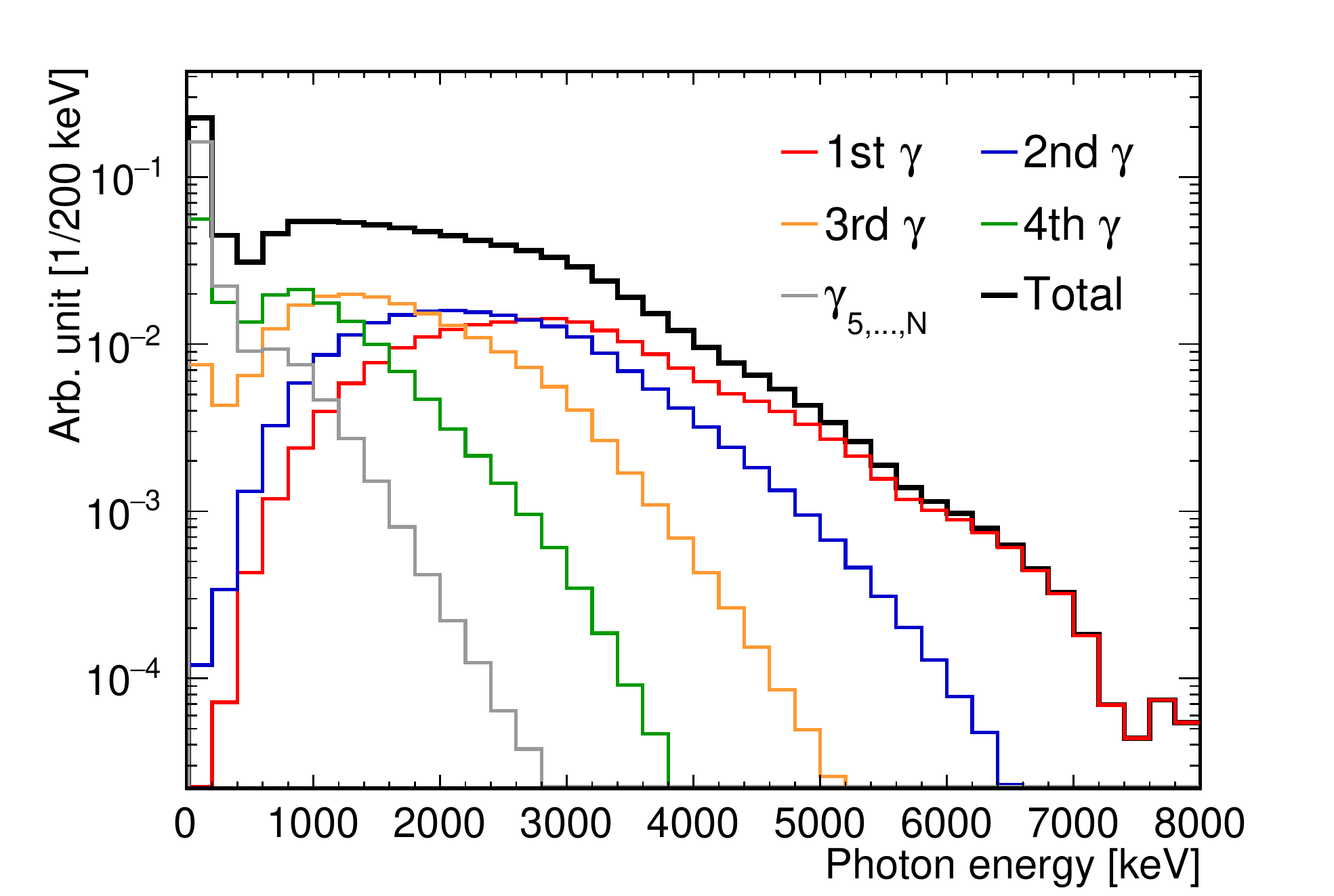}
  \end{minipage}%
  ~
  \begin{minipage}{0.49\textwidth}
  \includegraphics[trim=0.1cm 0.1cm 1.0cm 0.1cm, clip=true, 
  width=\textwidth]{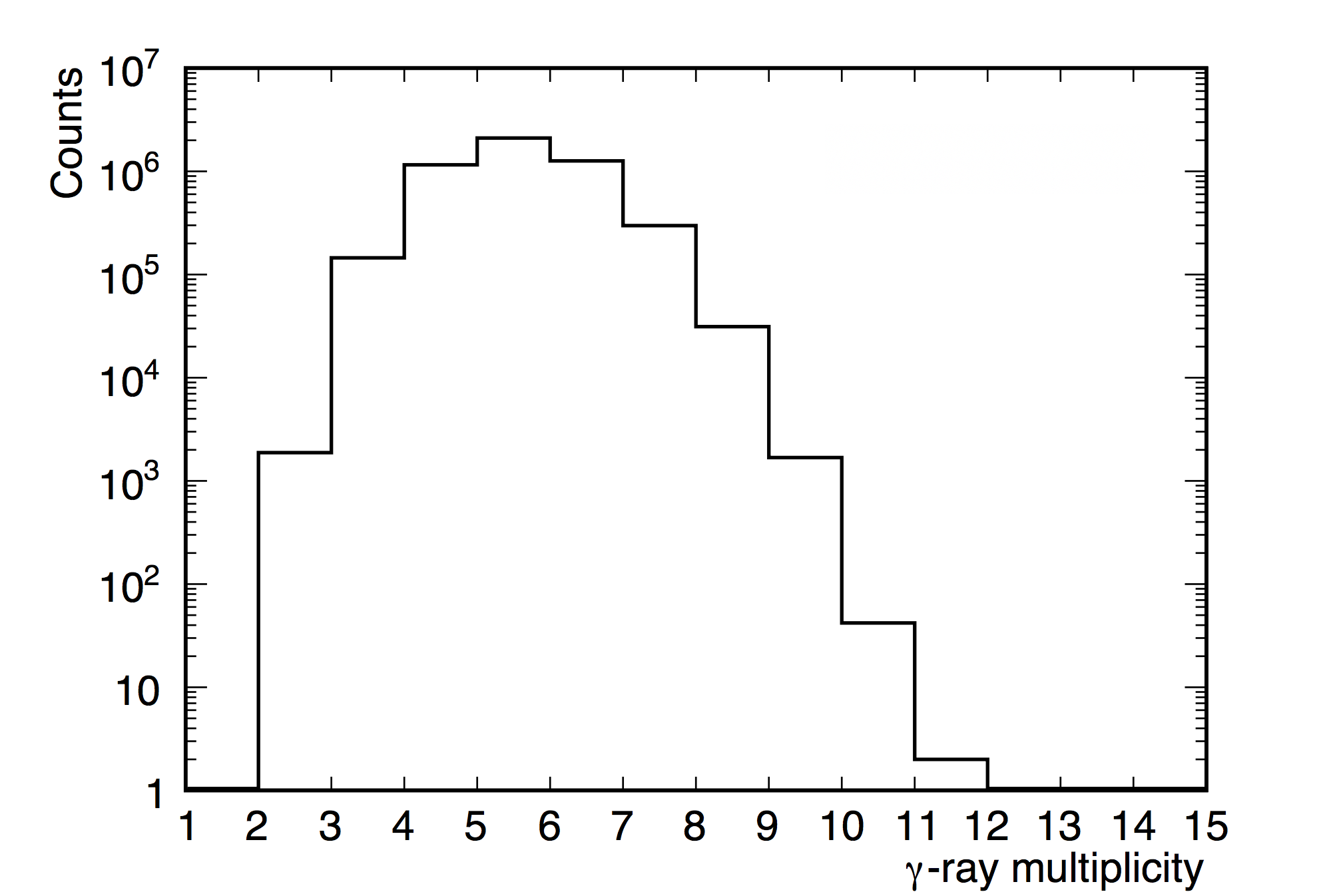}
  \end{minipage}%
  \caption{\textbf{Left:} Continuum component (black) according to our model for the 
           $\gamma$-ray energy spectrum from the thermal ${}^{157}$Gd$(n,\gamma){}^{158}$Gd reaction 
           and its composition in terms of contributions 
         from the first (red), second (blue), third (orange), fourth (green) $\gamma$ ray and
         other $\gamma$ rays (gray). The distributions are normalized such that the total
         continuum spectrum actually is a binned probability distribution and that the 
         relative contributions of the single components are properly reflected.  \textbf{Right:} 
          $\gamma$-ray multiplicity distribution obtained from 
         the continuum part of our spectrum model. Five million events were generated for the distributions.}
  \label{fig:Gd158_PDD}
\end{figure}

The PSF describes the coupling of a photon with given energy and multipolarity to the 
excited nucleus. Electric dipole (E1), magnetic dipole (M1) and electric quadrupole (E2) 
radiation are the most relevant multipolaries~\cite{Capote2009:RIPL3}. 
The experimental photonuclear data~\cite{Dietrich1988:Photoabs, Berman1975:Photoabs} show that the photoabsorption cross section 
of statically deformed spheroidal nuclei like ${}^{158}$Gd is well approximated 
by that of the Giant Dipole Resonance (GDR)  as the superposition of two Lorenzian lines, corresponding to oscilations
along each of the axes of the spheroid. 
The simplest model for PSF is thus called  the Standard 
Lorentzian Model (SLO)  having two Lorentzian forms~\cite{Axel1962:LorMdl, Capote2009:RIPL3}:
\begin{equation}
f_{\mathrm{E1}}^{\mathrm{(SLO)}}(E_\gamma) = 8.674 \cdot 10^{-8} \, \mathrm{mb}^{-1} \, 
\mathrm{MeV}^{-2}
\times \sum_i \frac{\sigma_i E_\gamma  \Gamma_i^2}{\left( 
E_\gamma^2 - E^2_i \right)^2 + E_\gamma^2 \Gamma^2_i } \, ,
\label{eq:E1_SLO}
\end{equation}
where usually two sets ($i=2$) of parameters are used to describe the two GDRs  in terms of a
resonance energy $E_i$ (in MeV), resonance width $\Gamma_i$ (in MeV) and a peak cross-section 
$\sigma_i$ (in mb). As shown in Fig.~\ref{fig:PSF}, this factor favors large $E_{\gamma}$ in the energy regime $<8$ MeV
and the factor $E_{\gamma}^{2L+1}$ favors large $E_{\gamma}$ as well. 
As a result of two competing factors of the nuclear  level density and the transmission coefficients, we obtain the broad peak 
structure in the continuum spectrum distribution of each $\gamma$ ray. 
The distribution $P(E_a,E_b)$ for the first, second, third and the fourth $\gamma$ ray 
(as in Fig.~\ref{fig:EmissionsModes}-left) is shown in Fig.~\ref{fig:Gd158_PDD}.
Their combination  for the total continuum spectrum is shown as well. The dips in the distributions at 0.4 MeV and 
7.5 MeV are due to a corresponding feature in the NLD model we employ.

Fig.\ref{fig:Gd158_PDD}-left shows a dominance of the continuum 
component above 5 MeV by the first $\gamma$ ray. 
This naturally suggests that the discrete peaks above 5 MeV are generated from the first
transition~\cite{Bollinger1970:NeutARC, Ali1994:Gd157}.  

For completeness, Fig.\ref{fig:Gd158_PDD}-right  shows the $\gamma$-ray multiplicity 
distribution as generated by the continuum part of our spectrum model. Some more remarks on this 
distribution are given in Sect.~\ref{subsubsec:Model:Model:Continuum}. 

\subsection {The MC Model (``ANNRI-Gd model'')}
\label{subsec:Model:Model}

In line with the GLG4sim approach~\cite{GLG4sim2005:HP}, our model for 
the $\gamma$-ray spectrum from the radiative 
thermal neutron capture reaction ${}^{157}$Gd$(n,\gamma)$ consists of two 
separate parts: the discrete peaks contribute $(6.94\pm 0.01)\%$, while 
the continuum component dominates the remaining $(93.06\pm 0.01)\%$ of  
our data. 

The  model  is written in C++ 
and is used through the program structure of 
our Geant4-based detector simulation.

\subsubsection {Continuum component}
\label{subsubsec:Model:Model:Continuum}


As already 
described in Sect.~\ref{subsec:Model:EmScheme}, we used the SLO model for the continuum 
since the E1 PSF dictates the general trend of the photon-nucleus coupling as a 
function of the $\gamma$-ray energy. To calculate $P(E_a,E_b)$ for E1 transitions with $E_\gamma = 
E_a - E_b$, we complete Eq.~(\ref{eq:continuum_prob_base}) with a proper normalization as
\begin{equation}
  P(E_a,E_b) = \frac{\mathrm{d}P}{\mathrm{d}E}(E_a,E_b) \, \delta E = 
  \frac{\rho(E_b) T_{E1}(E_\gamma)}{\int_{0}^{E_a} \rho(E'_b) T_{E1}(E'_\gamma)  \, 
  \mathrm{dE}'_b} \, \delta E \, , \quad  E'_\gamma = E_a - E'_b\, ,
\label{eq:continuum_prob}
\end{equation}
where $\delta E$ is a finite energy step in our computations. The E1 transmission coefficient is
\begin{equation}
  T_{E1}(E_\gamma) = 2 \pi \, E^3_\gamma \, f_{E1}(E_\gamma) \, .
\label{eq:E1_Trans}
\end{equation}

Note that, due to the normalization in Eq.~(\ref{eq:continuum_prob}), the absolute values of the NLD 
and the PSF do not matter for our purpose. 
The detailed comparison of our model with our data is presented
 in Sect.~\ref{sec:Performance}.

Fig.~\ref{fig:Gd158_PDD}-left shows the binned probability distribution $P(E_a,E_b)$ of the 
$\gamma$-ray energies generated by our continuum model part. It also shows the contributions of 
$\gamma$ rays from different transition steps in a cascade. 
In Fig.~\ref{fig:Gd158_PDD}-right one can see the corresponding $\gamma$-ray multiplicity distribution. 
The probability distributions $P(E_a,E_b)$ 
are tabulated for each energy $E_a$ for the continuum part. 
To prevent the 
generation of infinite cascades or small, negative $\gamma$-ray energies, we artificially force a 
cascade to end after a finite number of steps: If the remaining excitation $E$ falls below a 
threshold value of $E_{\mathrm{thr}}= 0.2$ MeV, one last $\gamma$ ray with low energy $E$ is 
generated. Therefore, one ``additional`` $\gamma$ ray is produced per cascade, effectively 
increasing the mean value by one. This procedure assures the total energy conservation.

\begin{figure}[b!t]
  \centering
  \includegraphics[trim=0.1cm 0.1cm 1.0cm 0.1cm, clip=true, 
  width=0.55\textwidth]{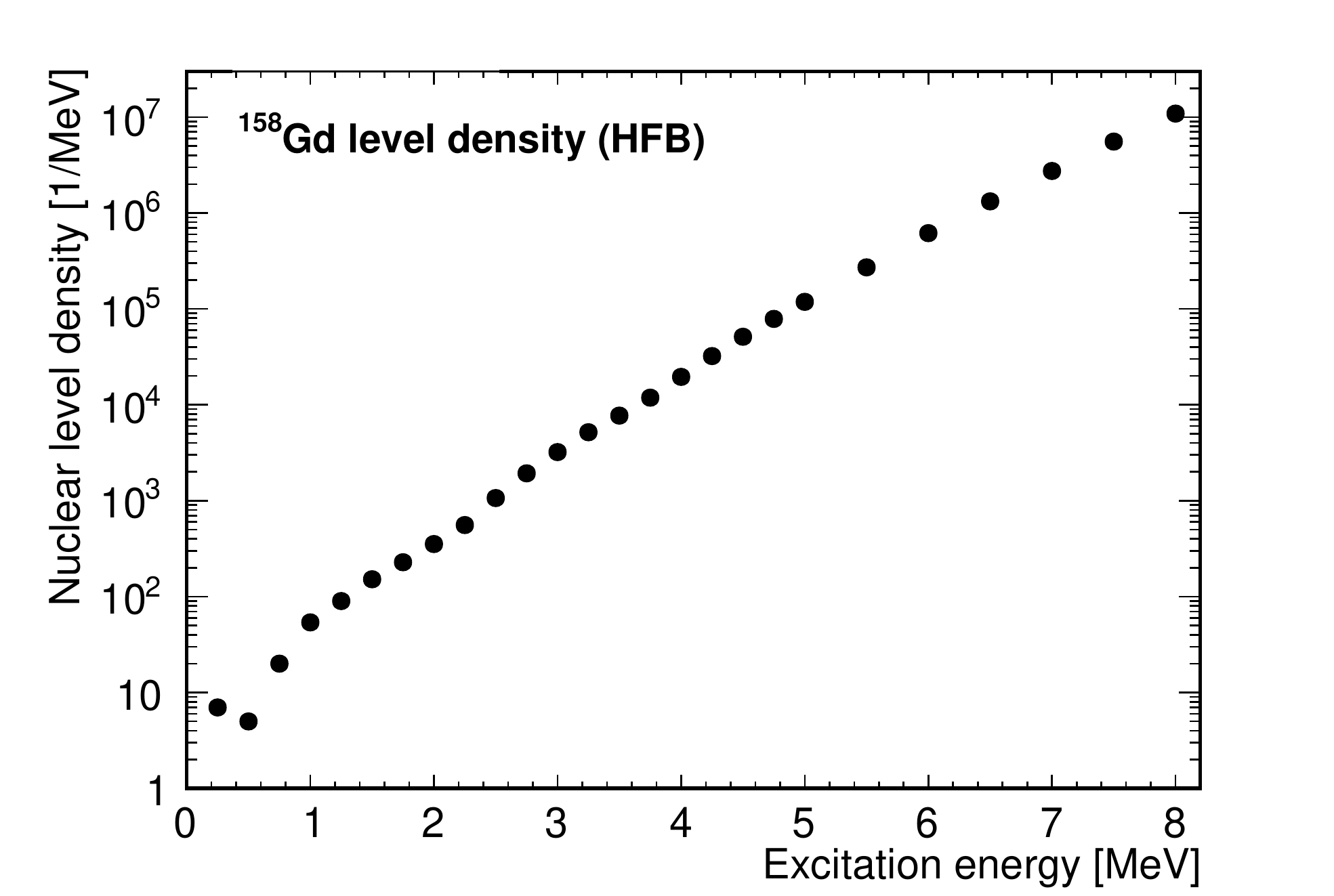}
  \caption{Tabulated values~\cite{Capote2009:RIPL3, RIPL32010:HP} for the NLD of ${}^{158}$Gd from 
           computations based on the HFB method~\cite{Goriely2007:HFB, Goriely2008:HFB}. 
           We used linear interpolation between the points, which have a 
           spacing of 0.25 MeV (0.5 MeV) below (above) 5 MeV excitation energy.}
  \label{fig:Gd158_NLD}
\end{figure}

\paragraph{Nuclear level density}
To describe the NLD of ${}^{158}$Gd as a function of excitation energy, we used a 
microscopic combinatorial level density computed according to the Hatree-Fock-Bogoliubov (HFB) 
method~\cite{Goriely2007:HFB, Goriely2008:HFB}. Tabulated values are provided separately for 
positive and negative parity levels~\cite{Capote2009:RIPL3, RIPL32010:HP}. We point-wise summed the 
two values and used linear interpolation in our calculations. For the modeling of the continuum 
component we used the HFB model results over the entire excitation energy range (Fig.~\ref{fig:Gd158_NLD}).

\paragraph{E1 photon strength function}

\begin{figure}[b!t]
  \centering
  \includegraphics[trim=0.1cm 0.1cm 1.0cm 0.1cm, clip=true, 
  width=0.55\textwidth]{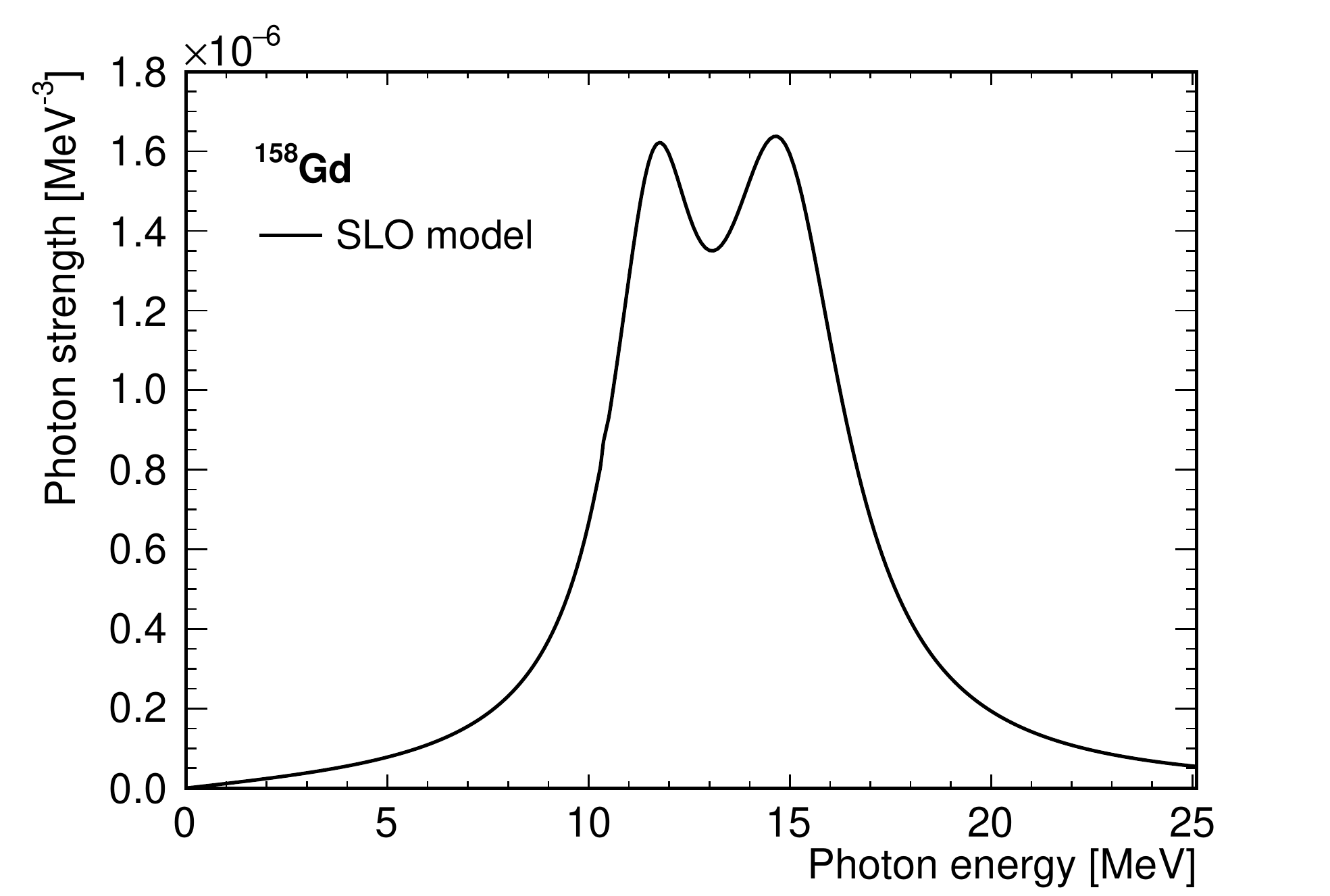}
  \caption{The E1 PSF for ${}^{158}$Gd, given as a function of the $\gamma$-ray 
         energy, that we used in our model for the spectral continuum component from the 
        thermal ${}^{157}$Gd$(n,\gamma)$ reaction. It follows the SLO model in 
        Eq.~(\ref{eq:E1_SLO}) with the parameters given in Table~\ref{tab:PSFParameters}.}
\label{fig:PSF}
\end{figure}

We list  four sets of values for the GDR parameters $E_i$, $\Gamma_i$ and $\sigma_i$  
in Table~\ref{tab:PSFParameters}~\cite{Shibata2011:JENDL4}. For our present  model,  only the first two 
E1 PSFs are used. 
Fig.~\ref{fig:PSF} shows the resulting PSF shape.
Two prominent GDR peaks $i=1,2$ are clearly visible. 
 

\begin{table}[t!b]
    \centering
    \caption{Parameter values for the PSF of the deformed ${}^{158}$Gd 
             nucleus~\cite{Shibata2011:JENDL4, JENDL4_HP,Vasilev1971:E1Params}.}
    \begin{tabular}{rccc}
        \hline
        Index $i$ & Cross-section $\sigma_i$ & Energy $E_i$ & Width $\Gamma_i$ \\
                  & [mb]                     & [MeV]        & [MeV]\\
        \hline
        1         & 249                     & 14.9        & 3.8 \\
        2         & 165                     & 11.7        & 2.6 \\
        3         & 3.0                     & 6.4         & 1.5 \\
        4         & 0.35                    & 3.1         & 1.0 \\ 
        \hline
    \end{tabular}
    \label{tab:PSFParameters}
\end{table}


\subsubsection {Discrete peaks}
\label{subsubsec:Model:Model:Peaks}

The previously described model for the continuum component of the $\gamma$-ray spectrum from the 
thermal ${}^{157}$Gd$(n,\gamma)$ relies on a continuous NLD description and 
thus does not reproduce sharp $\gamma$-ray energy lines in the observed spectra.
 We separately added this spectral component on top 
of the continuum part and tuned the strengths of the discrete peaks to match our measurement 
results.

\begin{figure}[b!t]
  \centering
  \begin{minipage}{0.49\textwidth}
  \includegraphics[trim=0.1cm 0.1cm 1.0cm 0.1cm, clip=true, 
  width=\textwidth]{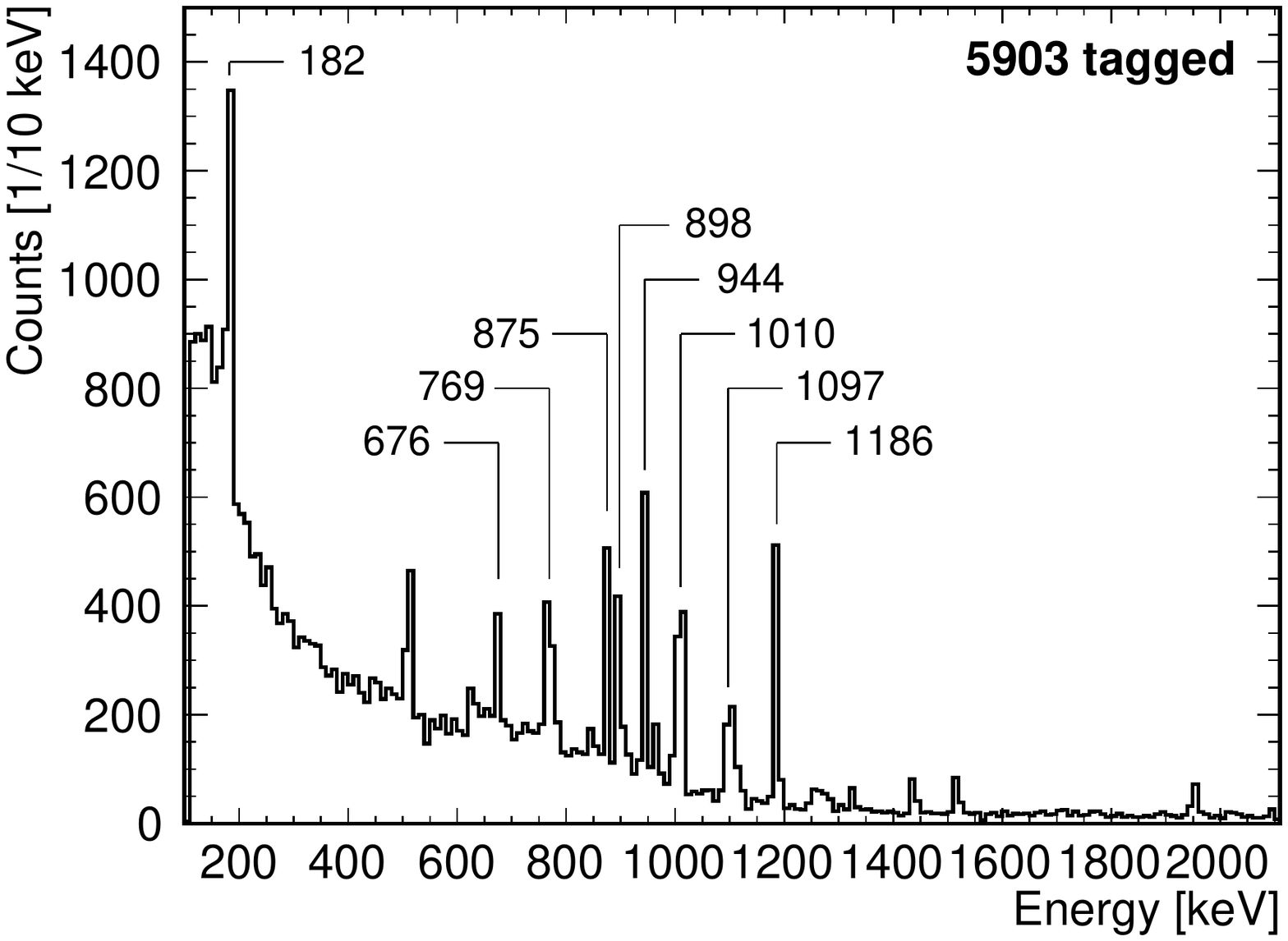}
  \end{minipage}%
  ~
  \begin{minipage}{0.49\textwidth}
  \includegraphics[trim=0.1cm 0.1cm 1.0cm 0.1cm, clip=true, 
  width=\textwidth]{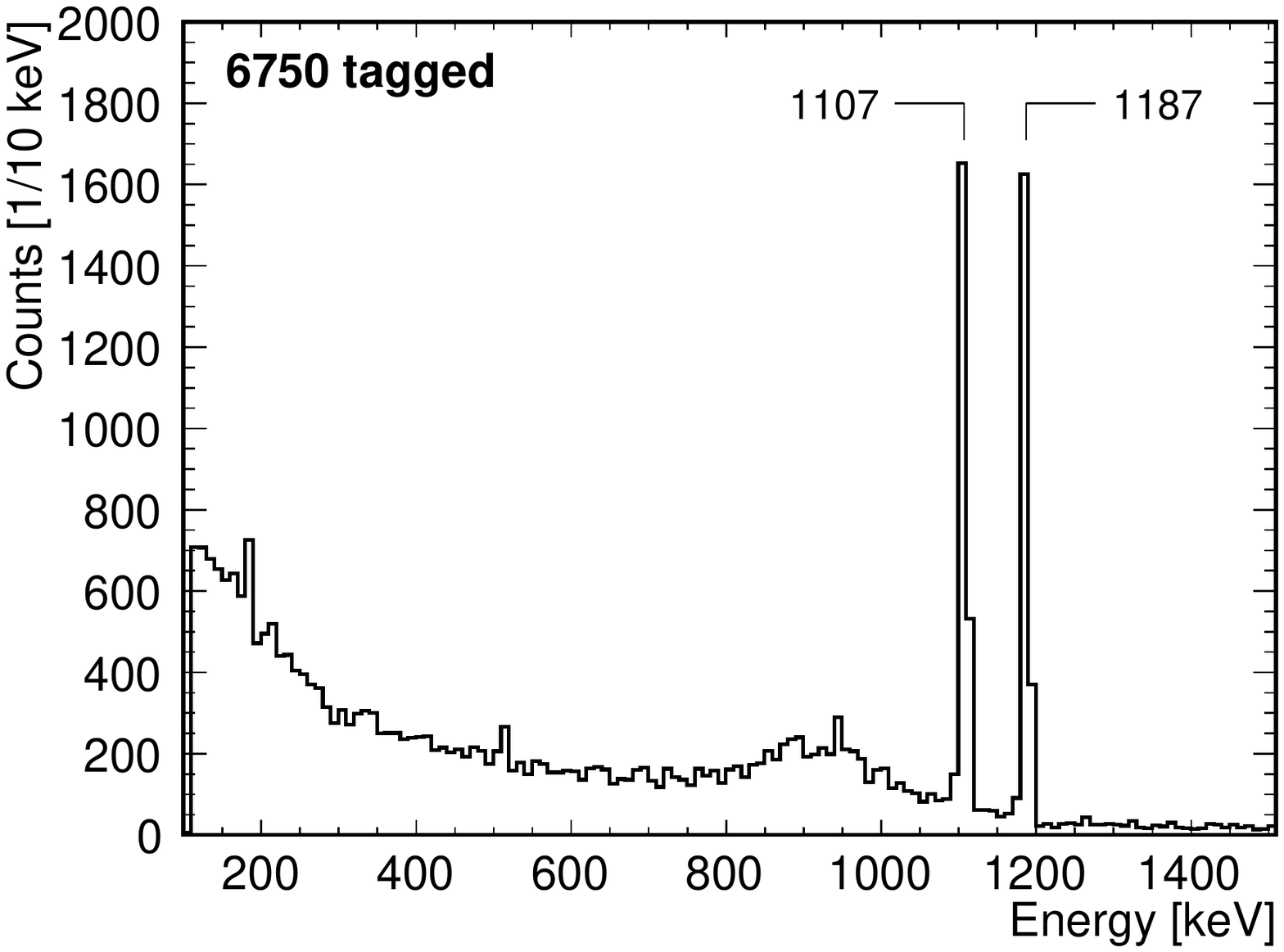}
  \end{minipage}%
  \caption{Combination of energy spectra from single crystals. The spectra show secondary 
           $\gamma$ rays observed in M2H2 events where the primary 5905 keV (left) or 6750 keV 
           (right) $\gamma$ ray was detected by another, non-neighboring crystal.}
  \label{fig:SecDiscPeaks}
\end{figure}

Using our data from all of the ANNRI detector's Ge crystals and selecting the dominant M1H1, 
M2H2 and M3H3 events, we identified 15 known~\cite{Nica2017:NuclDataGd158} 
discrete $\gamma$-ray lines above 5 MeV with high intensity after careful exclusion of single and 
double escape peaks. Following the previous assumption that the peaks in 
the high-energy part of the spectrum arise from the first transition, we refer to the corresponding 
$\gamma$ rays as `primary' $\gamma$ rays. We also identified $\gamma$ rays from subsequent transitions 
(`secondary' $\gamma$ rays) in detected multi-$\gamma$ events ($\mathrm{M}>1$) by looking at 
observed $\gamma$-ray energies besides the primary one used to tag the event. Two examples for the 
primary $\gamma$ rays (5903 keV and 6750 keV) are shown in Fig.~\ref{fig:SecDiscPeaks}. The 
energies of the identified primary and secondary transition $\gamma$ rays as well as their relative 
intensities as obtained from our data are listed in Table~\ref{tab:discPeaks}. 
Note that the direct transition from the neutron capture state ($J^\pi = 2^-$) to the ${}^{158}$Gd 
ground state ($J^\pi = 0^+$) is much suppressed because it is of M2 type. The energies in 
the table were not determined from our data but taken from~\cite{Nica2017:NuclDataGd158}. For 
cases where peaks obviously overlapped and could not be disentangled, we treated them combined in 
the intensity evaluation and list the mean primary $\gamma$-ray energy.

\begin{table}[t!b]
  \centering
  \caption{List of the 15 discrete peaks from primary $\gamma$ rays we identified in our data. The 
          stated energies are taken from Ref.~\cite{Nica2017:NuclDataGd158}, rounded to nearest keV. In four 
          cases the table lists the unweighted mean energy of known peaks that overlap in 
          our data:
          (i)   6001 keV combining 5995 keV and 6006 keV, 
          (ii)  5669 keV combining 5661 keV and 5677 keV, 
          (iii) 5595 keV combining 5582 keV, 5593 keV and 5.610 keV, as 
                well as
          (iv)  5167 keV combining 5155 keV and 5199 keV.
          (Intensities listed w.r.t. total data)}
  \label{tab:discPeaks}
  \begin{tabular}{r|c|c|c|c}
    \hline
    \multicolumn{4}{c|}{$\gamma$-ray energy [keV]} & Relative intensity\\
    \cline{1-4}
    Primary & \multicolumn{3}{|c|}{Secondary} & [$\times10^{-2} \%$]\\ 
    \hline 
    7937 & --  & -- & -- &  0.55 $\pm$ 0.03 \\ 
    7857 & --  & -- & -- &  2.38  $\pm$ 0.05  \\
    6960 & --  & -- & -- &  2.05  $\pm$ 0.06  \\
    6914 & 944 & -- & -- & 12.7  $\pm$ 0.1    \\ \hline
    \multirow{2}{*}{6750} & 1187 & --   & --  & 121   $\pm$ 3   \\ \cline{2-5}
                          & 1107 & --   & --  & 120   $\pm$ 3   \\ \hline
    \multirow{2}{*}{6672} & 1187 & --   & --  &  16   $\pm$ 1   \\  \cline{2-5}
                          & 1004 & 182  & --  &   2.9 $\pm$ 0.5 \\ \hline
    \multirow{3}{*}{6420} & 1517 & --   & --  &  12   $\pm$ 1   \\  \cline{2-5}
                          & 1438 & --   & --  &  14   $\pm$ 1   \\  \cline{2-5}
                          & 1256 & 182  & --  &   6.6 $\pm$ 0.8 \\ \hline
    \multirow{2}{*}{6001} & \multirow{2}{*}{749}  & 1187 & -- & 7.9  $\pm$ 0.2  \\  \cline{3-5}
				&   & 1107 & -- & 7.9  $\pm$ 0.2  \\ \hline
    \multirow{6}{*}{5903} & 1010 & 944  & --  &  46   $\pm$ 2   \\  \cline{2-5}
                          & 875  & 898  & 182 &  30   $\pm$ 2   \\   \cline{2-5}
                          &  \multirow{2}{*}{769}  & 1186 & --  &  22   $\pm$ 2   \\ \cline{3-5}
			&			   & 1004 & 182  &  4.0   $\pm$ 0.7   \\   \cline{2-5}
                          &  \multirow{2}{*}{676}  & 1279 & -- &  3.8   $\pm$ 0.5   \\  \cline{3-5}
			&			  & 1097 & 182 &  12   $\pm$ 1   \\ \hline
    5784 & 2073  & -- & -- & 19.7  $\pm$ 0.2  \\ \hline
    5669 & 2188  & -- & -- & 63.4    $\pm$ 0.3    \\ \hline
    5595 & 2262  & -- & -- & 66.7    $\pm$ 0.3  \\  \hline
    5543 & 2314  & -- & -- & 23.8  $\pm$ 0.2  \\ \hline
    5436 & 2421  & -- & -- & 16.2  $\pm$ 0.2  \\ \hline
    5167 & 2690  & -- & -- & 60.3    $\pm$ 0.3    \\ \hline
  \end{tabular}
\end{table}

The secondary $\gamma$ rays in Table~\ref{tab:discPeaks} were used as tag to identify 
further parts of the corresponding decay branches with information 
from Ref.~\cite{Nica2017:NuclDataGd158}. 

\begin{figure}[b!t]
  \centering
  \begin{minipage}{0.49\textwidth}
  \includegraphics[trim=0.1cm 0.1cm 1.0cm 0.1cm, clip=true, 
  width=\textwidth]{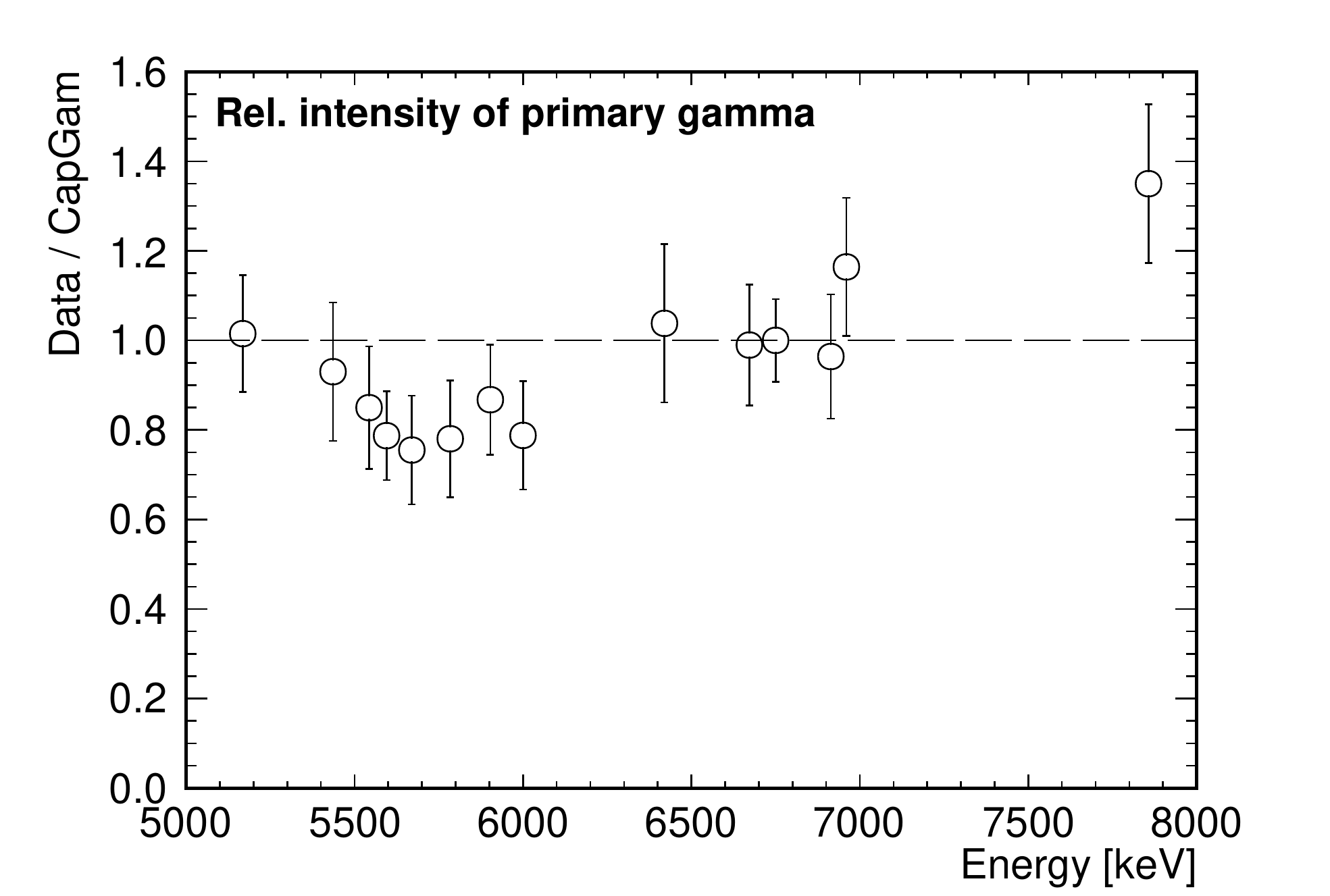}
  \end{minipage}%
  ~
  \begin{minipage}{0.49\textwidth}
  \includegraphics[trim=0.1cm 0.1cm 1.0cm 0.1cm, clip=true, 
  width=\textwidth]{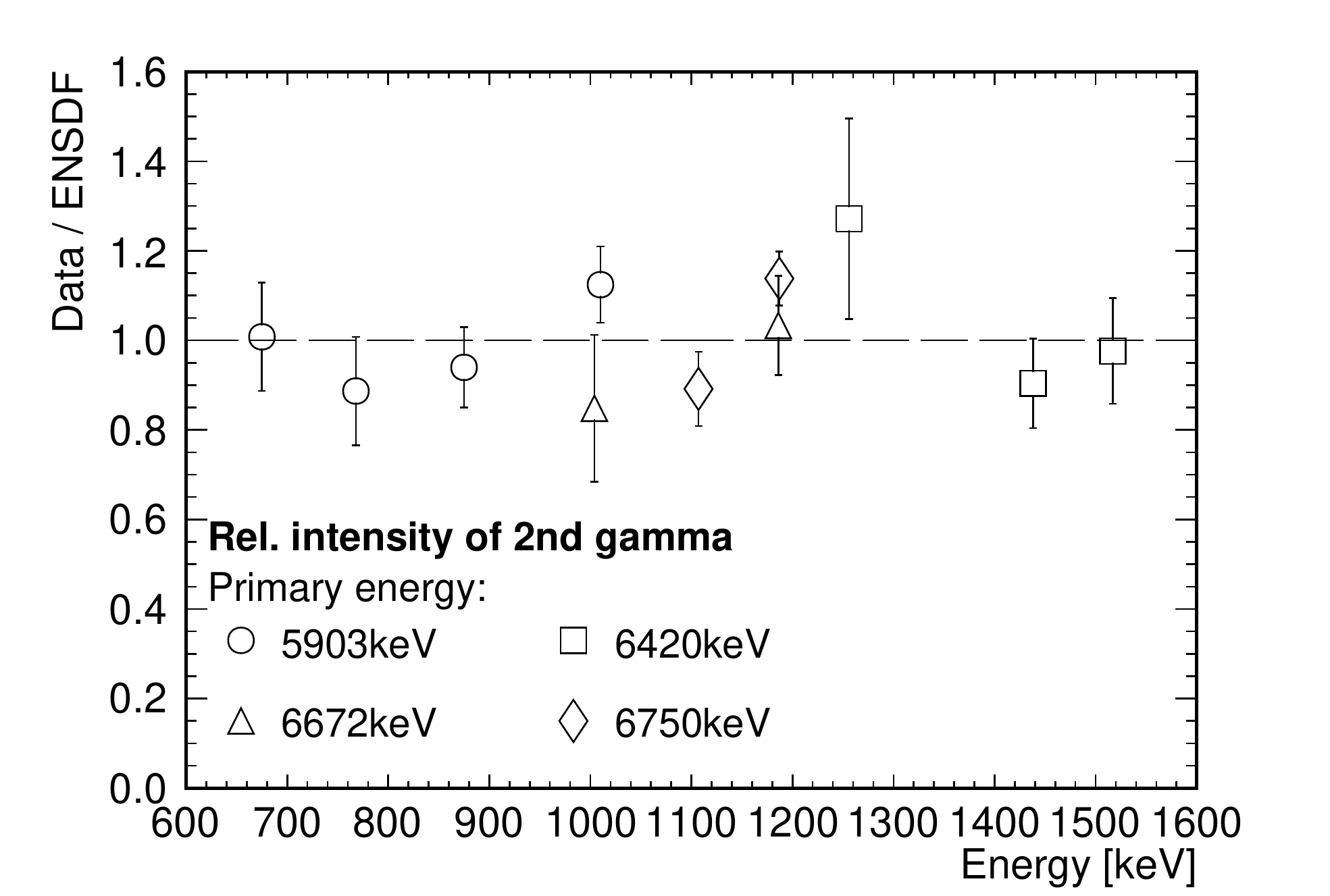}
  \end{minipage}%
  \caption{Ratios of our intensities for the primary (left) and the first secondary (right) 
           $\gamma$ rays in Table~\ref{tab:discPeaks} to the corresponding values listed in the 
           CapGam
           data base~\cite{CapGam_HP} and ENSDF~\cite{Nica2017:NuclDataGd158}.}
  \label{fig:DataVsCapGam}
\end{figure}

A comparison between the mean intensities from our data and values documented 
in Ref.~\cite{Nica2017:NuclDataGd158} is shown in 
Fig.~\ref{fig:DataVsCapGam} for primary $\gamma$ rays (left) and the secondary $\gamma$ rays 
(right) listed in Table~\ref{tab:discPeaks}. 
 
By summing all the relative intensities in Table~\ref{tab:discPeaks} we get $(6.94\pm 0.01)\%$ as 
an estimate for the fraction of neutron capture events that contribute to the discrete peaks. The 
remaining part, $(93.06\pm 0.01)\%$, contributes to the continuum part of the $\gamma$-ray spectrum.
We note that although our mean 
values for the discrete primary (secondary) $\gamma$-ray intensities in Table 7 are lower by about 20\% than the 
values in the literature, the effect of this difference on the total spectra is very small since the contribution of the discrete component 
to the total spectra is less than 7\%. 

\begin{figure}[h]
  \begin{center}
    \includegraphics[trim=0.1cm 0.1cm 1.0cm 0.1cm, clip=true,width=0.55\textwidth] 
    {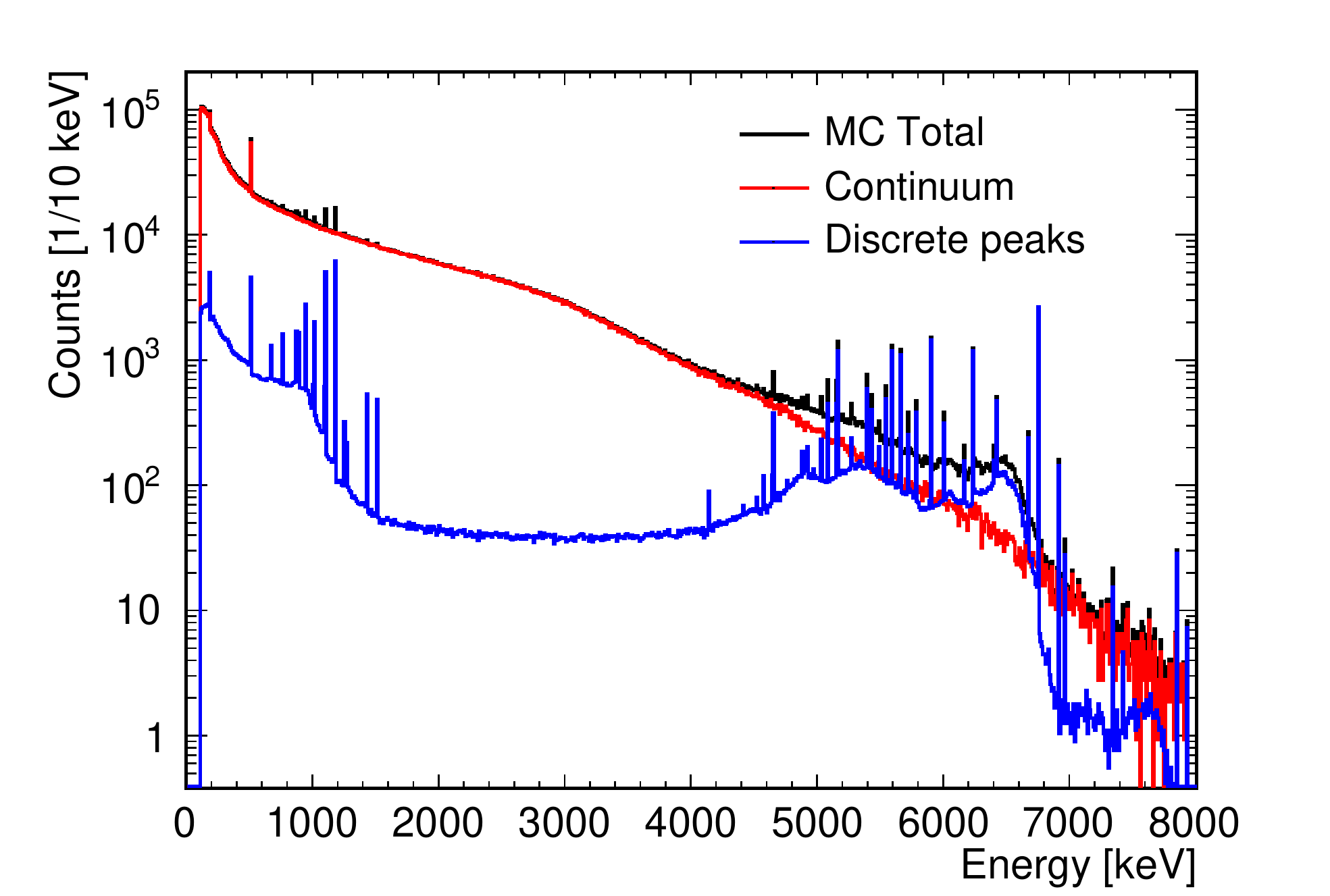}
    \caption{The MC-generated total energy spectrum from thermal ${}^{157}$Gd$(n,\gamma)$ for the peripheral crystal 6 (black), and the comprising continuum (red) and discrete (blue) component of the MC energy spectra shown separately.
             }
    \label{fig:ContDiscrete}
  \end{center}
\end{figure}

In order to implement the identified discrete peaks into our model, we converted the listed mean 
intensity values to probabilities summing to one and hard-coded them together with the $\gamma$-ray 
energies of the different cascades into our $\gamma$-ray generator. 
A particular cascade from 
Table~\ref{tab:discPeaks} is then generated according to this probability distribution. We finally obtain 
MC $\gamma$-ray spectrum as the sum of the continuum and discrete peaks as shown in Fig.~\ref{fig:ContDiscrete}. 

\section{Model performance}
\label{sec:Performance}

\begin{figure}[b!t]
  \centering
  \begin{minipage}{0.49\textwidth}
  \includegraphics[trim=0.1cm 0.1cm 1.0cm 0.1cm, clip=true, 
  width=\textwidth]{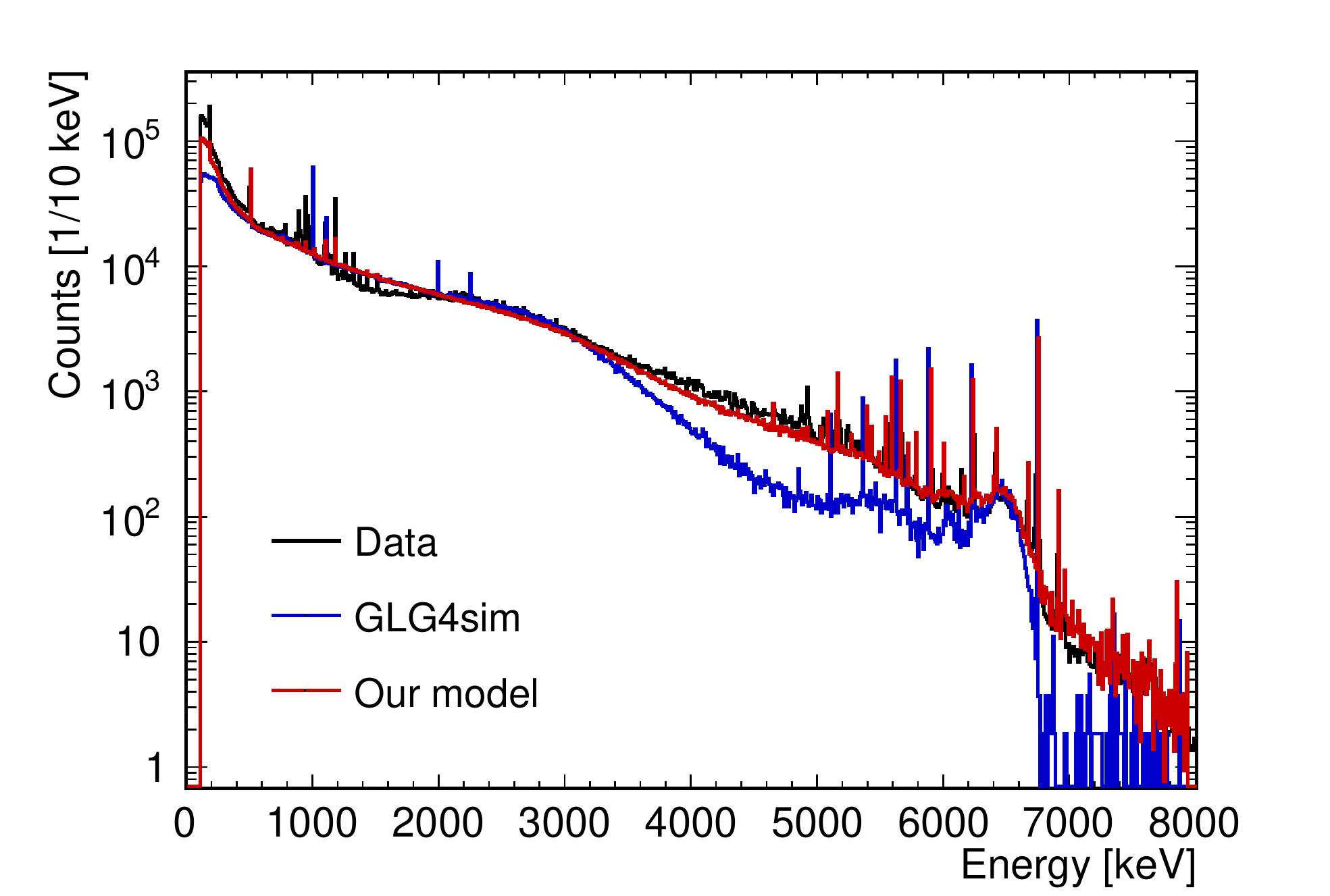}
  \end{minipage}%
  ~
  \begin{minipage}{0.49\textwidth}
  \includegraphics[trim=0.1cm 0.1cm 1.0cm 0.1cm, clip=true, 
  width=\textwidth]{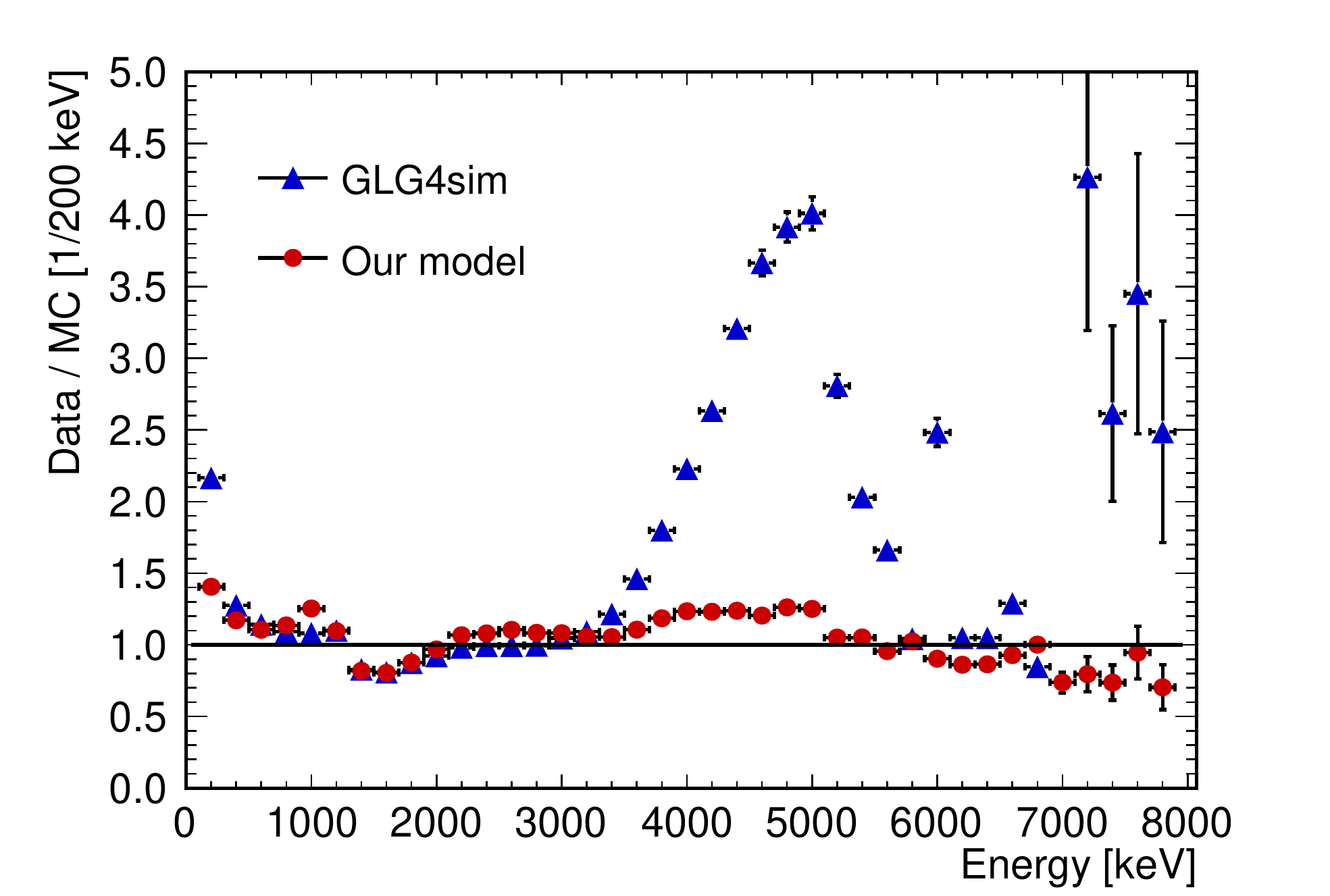}
  \end{minipage}%
  \caption{\textbf{Left:} Energy spectrum observed by the peripheral crystal 6 of the upper cluster 
           in M1H1 events from our measurement data (black), from the simulation with our 
           model (red) and from the simulation with the GLG4sim package (blue). 
           \textbf{Right:} Ratio between data and MC from the left side per 200 keV of observed 
           energy.}
  \label{fig:DataVsMc}
\end{figure}

To assess the performance of our model for the $\gamma$-ray spectrum from the thermal 
${}^{157}$Gd$(n,\gamma)$ reaction, we compared its output to the measured data.
With our model, we 
separately simulated $2\times10^8$ n-capture events for the discrete peaks and the continuum component,
and then merged in corresponding proportion. The resulting spectrum is shown in Fig.~\ref{fig:DataVsMc}, 
along with the data. 

On the left side of Fig.~\ref{fig:DataVsMc} one can see the resulting energy spectra observed by one of the 
 crystal for M1H1 events in our data and the simulations. The same for GLG4sim simulations is also shown.
Both simulated spectra were 
normalized such that the total sum of counts in each spectrum's range from 0.8 to 8 MeV is equal to 
the corresponding sum in the data spectrum. The low-energy limit excludes strong shape deviations 
below 0.8 MeV due to an excess of low-energy $\gamma$ rays in data. 
The shape of the energy spectrum in our data is 
significantly better reproduced by our  model, as seen from the ratios 
``Data/MC'' on the right of Fig.~\ref{fig:DataVsMc}. For the presented spectrum with 
the 200 keV binning, the mean deviation of the single ratios from the mean ratio is about 17\% for 
our model. We checked that the results for our model shown in Fig.~\ref{fig:DataVsMc} are consistent among the 
14 Ge crystals for different event classes. 

We also simulated the $\gamma$-ray spectrum with the generic photon evaporation model \cite{Allison:2016lfl} from the 
standard tools of the Geant4 framework. 
Resulting shape deviations with respect to the data were found to be larger than for the GLG4sim 
model.

The comparison between our data and our model for M1H1 events in Fig.~\ref{fig:DataVsMc}
shows some systematic discrepancies: From 0 to 1.2 MeV and from 2.2 to about 6 MeV the model 
underpredicts the number of $\gamma$ rays. Between 1.2 and 2.2 MeV and above 6 MeV it makes 
overpredictions. The shape description of these regions is interconnected because of the 
conservation of the energy for every capture reaction.

Overall, our  model significantly improves the description of the shape of the $\gamma$-ray 
energy spectrum from the thermal neutron capture on ${}^{157}$Gd with respect to other 
available models usable with the Geant4 structure. 

A second aspect to evaluate is the detected $\gamma$-ray multiplicity. 
Our experiment allows to classify events in 
terms of a multiplicity M of sub-clusters of hit crystals and the total number H of hit crystals. 
Table \ref{tab:GdMultiplicities} lists the fractions of events from the classes 
$\mathrm{M}=1,2,3,4$, $\mathrm{M} \leq \mathrm{H} \leq 4$ in our data and in the MC sample from our 
 model.


\begin{table}[t!b]
    \centering
    \caption{Fractions of events from the classes with $\mathrm{M}=1,2,3,4$ and $\mathrm{M} \leq 
             \mathrm{H} \leq 4$ in our data (Exp) and the MC sample from our model. The 
             column `Ratio' lists the ratio of the experimental value and the MC value. In the 
             column `Total', the upper value is the sum of the experimental values, the middle
             value is the sum of the MC values and the lower value in square brackets is the ratio
             of the above values. All errors are from statistics only.}
    \label{tab:GdMultiplicities}
    \begin{tabular}{c|c||c||cc|c}
    
      \hline 
      \multicolumn{2}{c||}{Class} & Data & Fraction [\%] & Ratio (Exp/MC)& Total\\
      \hline
      M & H &  &  & & \\
      \hline \hline
      \multirow{8}{*}{1} & \multirow{2}{*}{1} & Exp & $69.936\pm0.008$ &   \multirow{2}{*}{$0.971 \pm 0.013$} & \\
                         &                    & MC  & $72\pm1$         & 					&\\
      \cline{2-5}
                         & \multirow{2}{*}{2} & Exp & $17.161\pm0.007$ &   \multirow{2}{*}{$1.015 \pm 0.018$} & \\
                         &                    & MC  & $16.9\pm0.3$     & 		&          $89.648\pm0.011$ \\
      \cline{2-5}
                         & \multirow{2}{*}{3} & Exp & $2.315\pm0.003$  &   \multirow{2}{*}{$0.897\pm 0.014$} & $91.8\pm1.0$\\
                         &                    & MC  & $2.58\pm0.04$    & & Ratio=$[0.977\pm0.011]$\\
      \cline{2-5}
                         & \multirow{2}{*}{4} & Exp & $(2.364\pm0.009)\times10^{-1}$ &  \multirow{2}{*}{$0.844\pm0.012$} & \\
                         &                    & MC  & $(2.80\pm0.04)\times10^{-1}$  & & \\
      \hline
      \multirow{6}{*}{2} & \multirow{2}{*}{2} & Exp & $7.401\pm0.005$  &   \multirow{2}{*}{$1.298\pm0.018$} & \\
                         &                    & MC  & $5.70\pm0.08$    & \\
      \cline{2-5}
                         & \multirow{2}{*}{3} & Exp & $2.056\pm0.003$ &    \multirow{2}{*}{$1.071\pm0.017$} & $9.858\pm0.006$\\
                         &                    & MC  & $1.92\pm0.03$    & & $8.0\pm 0.1$ \\
      \cline{2-5}
                         & \multirow{2}{*}{4} & Exp & $(4.01\pm0.01)\times10^{-1}$ & \multirow{2}{*}{$0.950\pm0.014$} & Ratio=$[1.226\pm 0.013]$\\
                         &                    & MC  & $(4.22\pm0.06)\times10^{-1}$ & & \\
      \hline
      \multirow{4}{*}{3} & \multirow{2}{*}{3} & Exp & $(2.96\pm0.01)\times10^{-1}$ & \multirow{2}{*}{$1.370\pm0.020$} & \\
                         &                    & MC  & $(2.16\pm0.03)\times10^{-1}$ & &    $(3.807\pm0.011)\times10^{-1}$ \\
      \cline{2-5}
                         & \multirow{2}{*}{4} & Exp & $(8.47\pm0.05)\times10^{-2}$ &  \multirow{2}{*}{$1.210 \pm 0.019$} & $(2.86\pm0.03)\times10^{-1}$ \\
                         &                    & MC  & $(7.0\pm0.1)\times10^{-2}$  & &          Ratio=$[1.331\pm0.015]$\\
      \hline
      \multirow{2}{*}{4} & \multirow{2}{*}{4} & Exp & $(5.9\pm0.1)\times10^{-3}$ &    \multirow{2}{*}{$1.64\pm0.05$} & same as\\
                         &                    & MC  & $(3.6\pm0.1)\times10^{-3}$ & & left \\
      \hline   
    \end{tabular}
\end{table}


The frequency of the most dominant event classes, M1H1 and M2H2, is well reproduced 
within about 3\%; the total agreement between MC and data for $M=1$ is about 2\%. For event classes 
with MxHx, $x \geq 2$, which have lower frequency / statistics, the MC tends towards increasing
underpredictions. 

\section{Conclusion}
\label{sec:Summary}

A good model for the $\gamma$-ray energy spectrum from the radiative thermal neutron capture on 
natural gadolinium is an important prerequisite for MC studies to evaluate the efficiency to tag 
neutrons from IBD events in gadolinium-enhanced $\overline{\nu}_e$ searches. This is especially true 
for water Cherenkov or segmented detectors, where the energy distribution within a $\gamma$-ray 
cascade from the neutron capture on gadolinium heavily impacts the detectability of this marker 
signal.

Using the Ge spectrometer of the ANNRI detector at MLF, J-PARC, we performed a 
measurement of the $\gamma$-ray energy spectrum from thermal neutron capture on ${}^{157}$Gd. 


Based on our data and a Geant4 simulation of the ANNRI detector, we have developed a model for the 
$\gamma$-ray energy spectrum from the thermal ${}^{157}$Gd$(n,\gamma){}^{158}$Gd reaction. This 
marks an important step towards a model for natural gadolinium, which is of use for 
gadolinium-enhanced $\overline{\nu}_e$ measurements.

While the 
strength information of 15 discrete peaks above 5 MeV in our data is directly included into the 
spectrum model, the continuum component is modeled using a statistical approach. 
We used the Standard Lorentzian PSF model to describe the 
strength of the photon--nucleus coupling as a function of the $\gamma$-ray energy. 

The measured spectrum agrees within $\sim$17\% with that of our ANNRI-Gd model at 200 keV binning.
We found this outcome to be a significant improvement compared 
to other spectrum descriptions, e.g., from the standard Geant4 or the GLG4sim package. 
The completeness of our model however lies in including the contribution from the thermal ${}^{155}$Gd$(n,\gamma){}^{156}$Gd reaction (the other prime component for the natural gadolinium), which we shall report soon.
%

\section*{Acknowledgement}
\label{sec:Acknowledgements}

This work is supported by the JSPS Grant-in-Aid for Scientific Research on Innovative Areas 
(Research in a proposed research area) No. 26104006. It benefited from the use of the 
neutron beam of the JSNS and the ANNRI detector at the Material and Life Science Experimental 
Facility of the Japan Proton Accelerator Research Complex.

\appendix{
\section*{Appendices}
\section{Efficiency Calculation}
\label{sec:AppEfficiency}

For radioactive sources / excited nuclei that can emit more than one $\gamma$ ray per 
decay ($^{60}$Co, $^{152}$Eu and $^{36}$Cl), a reduction of the photopeak efficiency due 
to the trigger / veto condition has to be taken into account: If one or more secondary 
$\gamma$ rays 
are emitted along with the primary $\gamma$ ray of energy $E_\gamma$, there is a chance that one of 
the secondary $\gamma$ rays vetoes the primary $\gamma$ ray hit by directly going into the BGO 
shield of the corresponding Ge  cluster. This effectively reduces the
photopeak efficiency compared to the case where solely the primary $\gamma$ ray would be emitted. 

The $\gamma$ rays from the thermal ${}^{35}$Cl$(n,\gamma){}^{36}$Cl reaction 
do not allow the determination of absolute efficiency values since the number of emitted 
$\gamma$ rays is unknown. Therefore, we computed efficiency values relative to 
the photopeak efficiency of the most intense line at 7414 keV among our selected lines. The 
normalization of the reference efficiency was obtained from our MC simulation.

We corrected the single photopeak efficiency for this trigger effect differently for $^{36}$Cl / $^{152}$Eu and $^{60}$Co. 
From the complex decay and deexcitation schemes of $^{36}$Cl and $^{152}$Eu we only selected 
$\gamma$ rays for the efficiency determination that are dominantly emitted alone or with just one 
additional $\gamma$ ray in their particular decay channel: 5517 keV, 7414 keV, 
7790 keV and 8579 keV for $^{36}$Cl; 344 keV, 779 keV, 1112 keV and 
1408 keV for $^{152}$Eu. Relevant branching ratios can be found 
in Ref.~\cite{Cameron2013:NuclDataA37, Martin2013:NuclDataA152}. This selection allowed for an easier 
estimation of the above described inefficiency in the two $\gamma$-ray cases by multiplying the raw 
photopeak efficiency value for a crystal with the correctition
\begin{equation}
 C_i = \frac{\varepsilon_i^{\mathrm{MC}}(E_\gamma)}{\varepsilon_{i,2\gamma}^{\mathrm{MC}}(E_\gamma; 
E_{\gamma2})}
\end{equation}
coming from our Geant4 MC simulation. It is calculated from the single photopeak MC efficiency 
$\varepsilon_i^{\mathrm{MC}}(E_\gamma)$ for the $\gamma$ ray of interest with energy $E_\gamma$ and 
the corresponding single photopeak MC efficiency $\varepsilon_{i,2\gamma}^{\mathrm{MC}}(E_\gamma; 
E_{\gamma2})$ obtained when the second $\gamma$ ray with $E_{\gamma2}$ is simultaneously propagated 
through the detector.

For the $^{60}$Co source, which essentially always emits two $\gamma$ rays ($E_1=1173$ keV 
and $E_2 = 1332$ keV)~\cite{Browne2013:NuclDataA60}, we determined the 
corrected single photopeak efficiency directly through a fit: We look at a pair of crystals $(i,j)$, $i 
\neq j$, where each crystal is on a separate cluster of ANNRI. The number of observed M1H1 events 
where $E_k$ ($k=1,2$) is deposited in crystal ($i$) is $N_{ik}$ with its error $\sigma_{ik}$. One 
expects this value to be $\overline{N}_{ik} = \beta T r_{L,i} \varepsilon_{ik} (1-C_i)$ with 
$\varepsilon_{ik} \equiv \varepsilon_{i}(E_k)$, the live time $T$ and the dead time correction 
factor $r_{L,i}$. The efficiency correction $(1-C_i)$ is due to the 
inefficiency described above. For a given pair of crystals and the two $\gamma$ rays this yields 
four combinations of crystal and $\gamma$ ray. Moreover, we look at the coinciding detection of 
both $\gamma$ rays in M2H2 events by the crystals $(i,j)$. With $E_l \neq E_k$ being the second 
$\gamma$ ray, the observed number of coincidence events where $E_k$ ($E_l$) is detected in crystal 
$i$ ($j$) is $N_{ikjl}$. Its error is $\sigma_{ikjl}$. The expected value is 
$\overline{N}_{ikjl} = \beta T r_{L,ij} \varepsilon_{ik}  \varepsilon_{jl} W(\theta_{ij})$. Here, 
$r_{L,ij}$ is the dead time correction factor for the crystal pair $(i,j)$, which typically is on 
the order of 90\%. The factor $W(\theta_{ij})$ accounts for the predicted 
angular correlation~\cite{Larsen1969:Co60AngCorr} of the $\gamma$ rays from $^{60}$Co with 
 angle $\theta_{ij}$, which is given by the angle of the detector pair 
$(i,j)$. A second combination, $N_{iljk}$, simply follows from permuting the $\gamma$ ray 
energies. With the in total six observables we minimized the expression
\begin{equation}
\begin{split}
 \chi^2_{ij} &= \left( \frac{N_{ik}   - \overline{N}_{ik}}  {\sigma_{ik}}   \right)^2 +
                \left( \frac{N_{il}   - \overline{N}_{il}}  {\sigma_{il}}   \right)^2 +
                \left( \frac{N_{jk}   - \overline{N}_{jk}}  {\sigma_{jk}}   \right)^2 + 
                \left( \frac{N_{jl}   - \overline{N}_{jl}}  {\sigma_{jl}}   \right)^2 \\&+
                \left( \frac{N_{ikjl} - \overline{N}_{ikjl}}{\sigma_{ikjl}} \right)^2 +
                \left( \frac{N_{iljk} - \overline{N}_{iljk}}{\sigma_{iljk}} \right)^2
\end{split}
\end{equation}
for 48 crystal pairs $(i,j)$, one was excluded, to fit the four uncorrected single photopeak 
efficiencies, $\varepsilon_{ik}$, $\varepsilon_{il}$, $\varepsilon_{jk}$ and $\varepsilon_{jl}$, and
$\beta T$ for different but fixed values of the constant $C$. The best agreement between the mean 
of the fitted values of $\beta T$ and the nominal value was obtained for $C=0.225$. Using this 
constant, we took the averages of the efficiency values per crystal and energy as final results.
With this method we obtained a single photopeak efficiency of $(1.3\pm 0.1)\%$ at 1.3 
MeV for all 14 Ge crystals combined.

Figure~\ref{fig:CombinedEff} depicts the ratios of the single photopeak efficiencies from data and 
from MC at the single $\gamma$ ray energies averaged over all 14 crystals (left) and for all 14 
crystals averaged over the 11 data points (right). On both plots one can see that the weighted mean 
values of the ratios deviate by less than 10\% from perfect agreement and maximum deviations 
are about 20\%. The weighted sample standard deviation of the ratios for all crystals and 
data points is about 6\%. From this study,  
we conclude that we understand the photopeak efficiency of 
each crystal not only over the energy range from 344 keV to 8579 keV but also uniformly 
over the entire solid angle of the detector  and that 
we can reproduce the responce of each crystal very well by our Geant4 detector 
simulation.

\begin{figure}[h]
  \centering
  \begin{minipage}{0.49\textwidth}
  \includegraphics[trim=0.1cm 0.1cm 1.0cm 0.1cm, clip=true, 
  width=\textwidth]{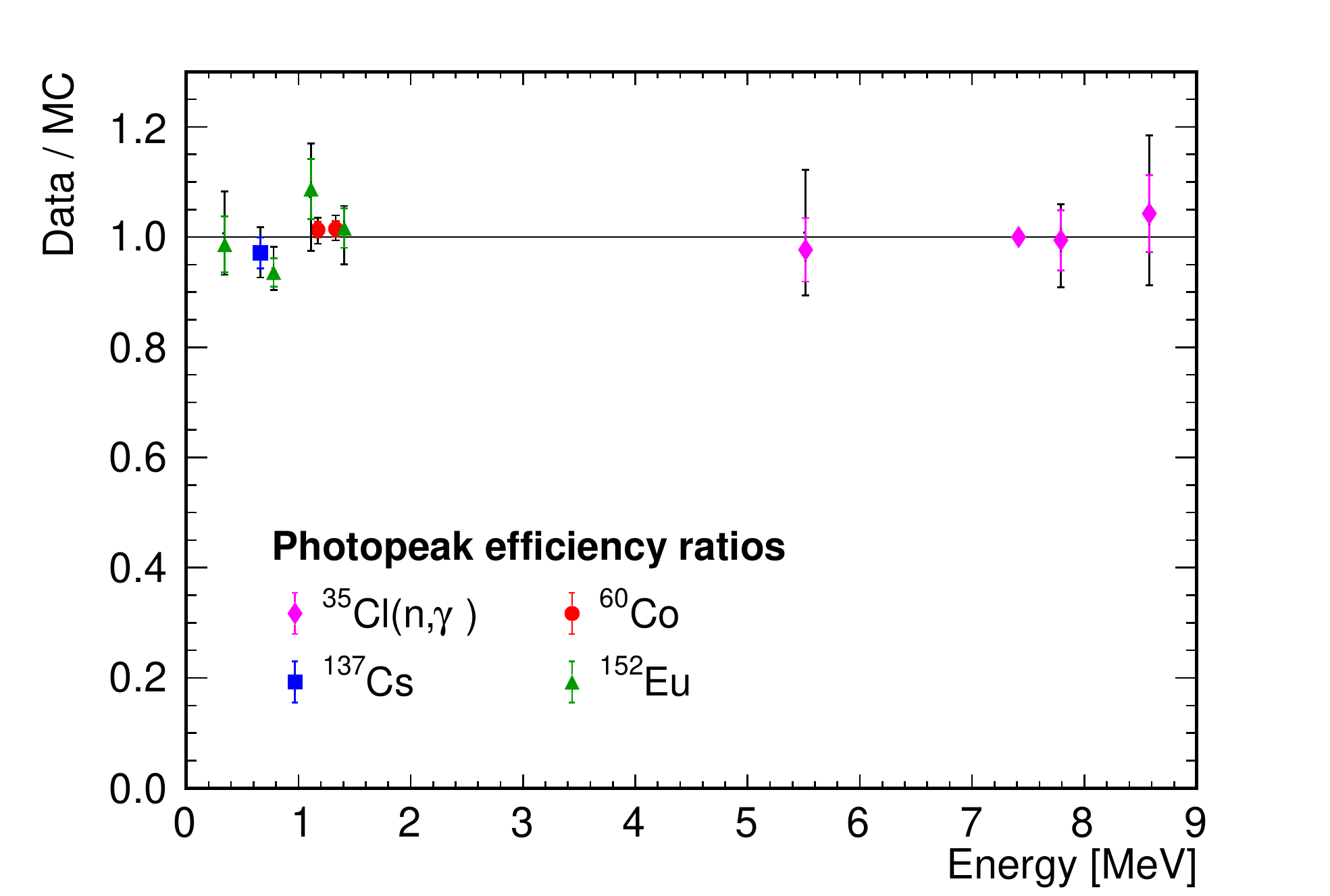}
  \end{minipage}%
  ~
  \begin{minipage}{0.49\textwidth}
  \includegraphics[trim=0.1cm 0.1cm 1.0cm 0.1cm, clip=true, 
  width=\textwidth]{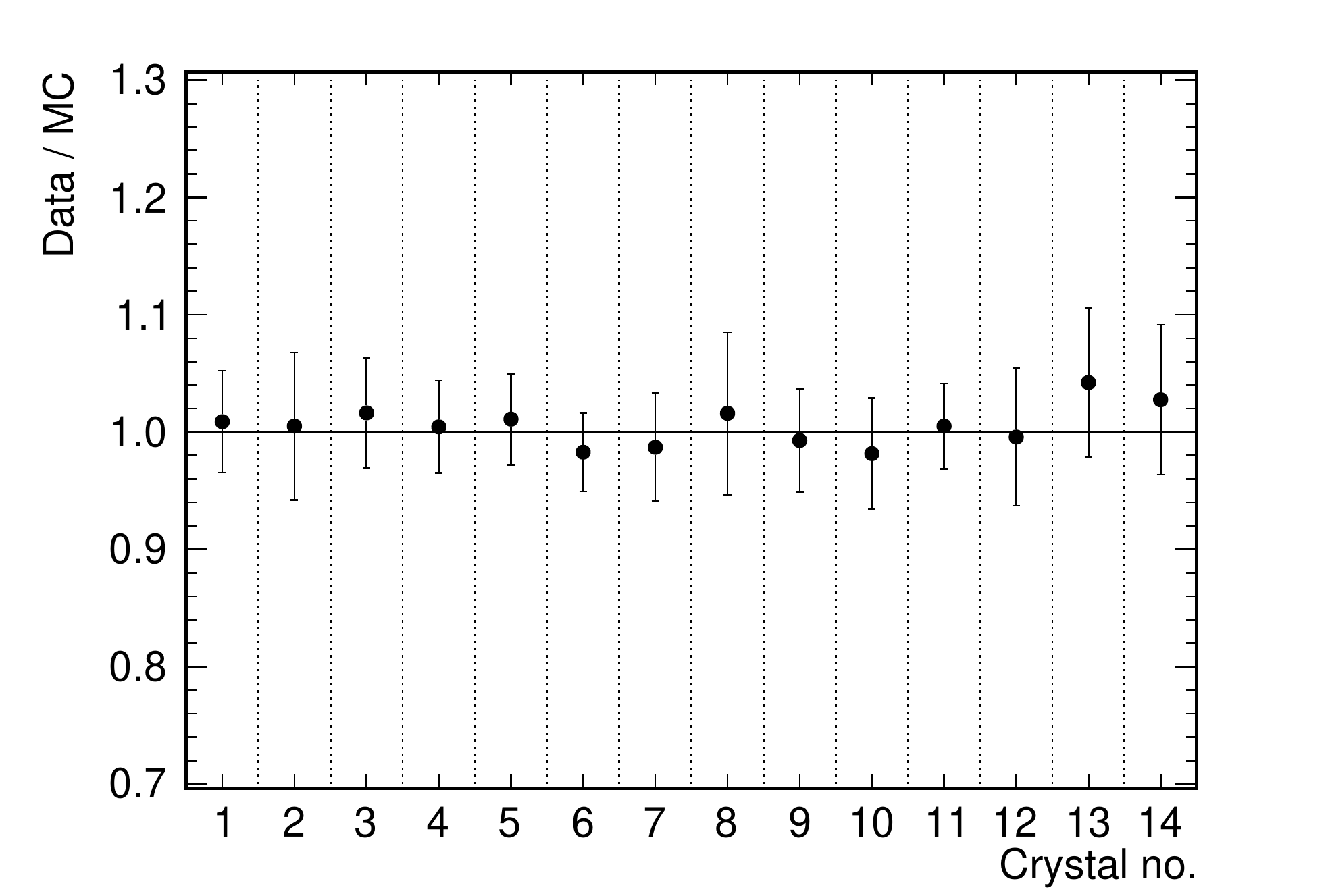}
  \end{minipage}%
  \caption{Ratios of the single photopeak efficiencies from data to single photopeak efficiencies 
           from MC (see Eq.~(\ref{eq:MCeff})) averaged over the 14 crystals at the fixed 
           $\gamma$-ray 
           energies (left) and averaged over the 11 data points for the single crystals (right).
           Linear interpolation between the points from the simulation was used to determine the MC
           efficiency at an intermediate energy. The calculations of the weighted
           mean values and the weighted sample standard deviations (error bars) take the 
           errors of the data points (see Fig.~\ref{fig:PhotopeakEff}) into account. 
           Outer error bars indicate the extreme values of the ratios in the respective samples. 
           The ${}^{35}$Cl$(n,\gamma)$ data point at 7414 keV is the reference for the 
           normalization of the other data points of this reaction. It perfectly agrees with a 
           ratio of one since it was normalized with the MC simulation.}
  \label{fig:CombinedEff}
\end{figure}

\section{Double/Triple Gamma Spectrum:}
\label{sec:m2h2m3h3}

The M2H2 and the M3H3 classified data are the next dominant fraction of the data after the M1H1 discussed above. The corresponding spectra in data also agree with those generated by our model as shown in Fig.~\ref{fig:mh2233}. With increased multiplicity, the slope of the spectrum grows softer, owing to reduced probability of emitting higher energy gamma rays, as also seen in Fig.~\ref{fig:multispectra} and independent checks from simulations as in the Fig.~\ref{fig:Gd158_PDD}.
\begin{figure}[h]
  \begin{center}
    \includegraphics[trim=0.1cm 0.1cm 1.0cm 0.1cm, clip=true,width=0.49\textwidth]
    {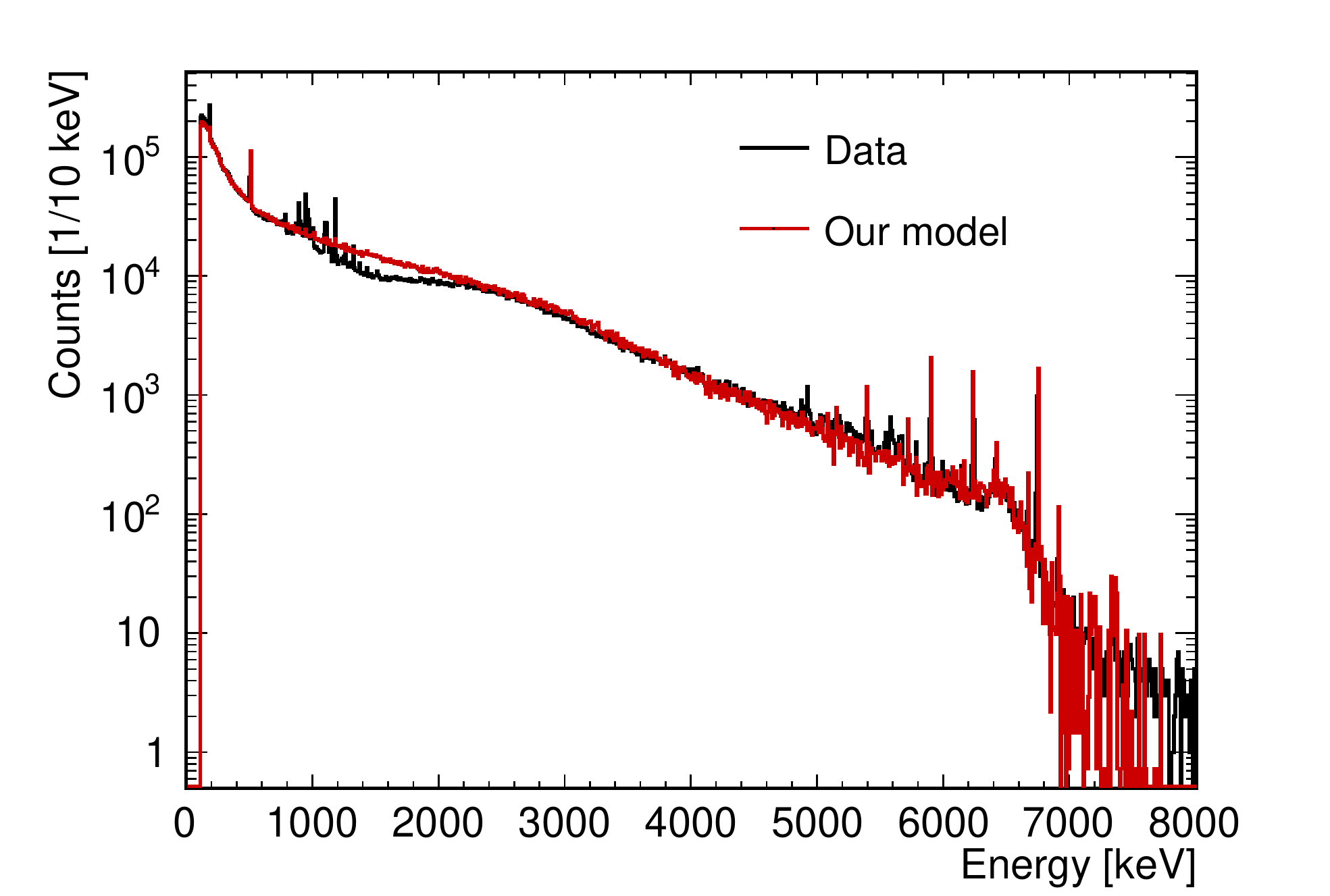}
    \includegraphics[trim=0.1cm 0.1cm 1.0cm 0.1cm, clip=true,width=0.49\textwidth]
    {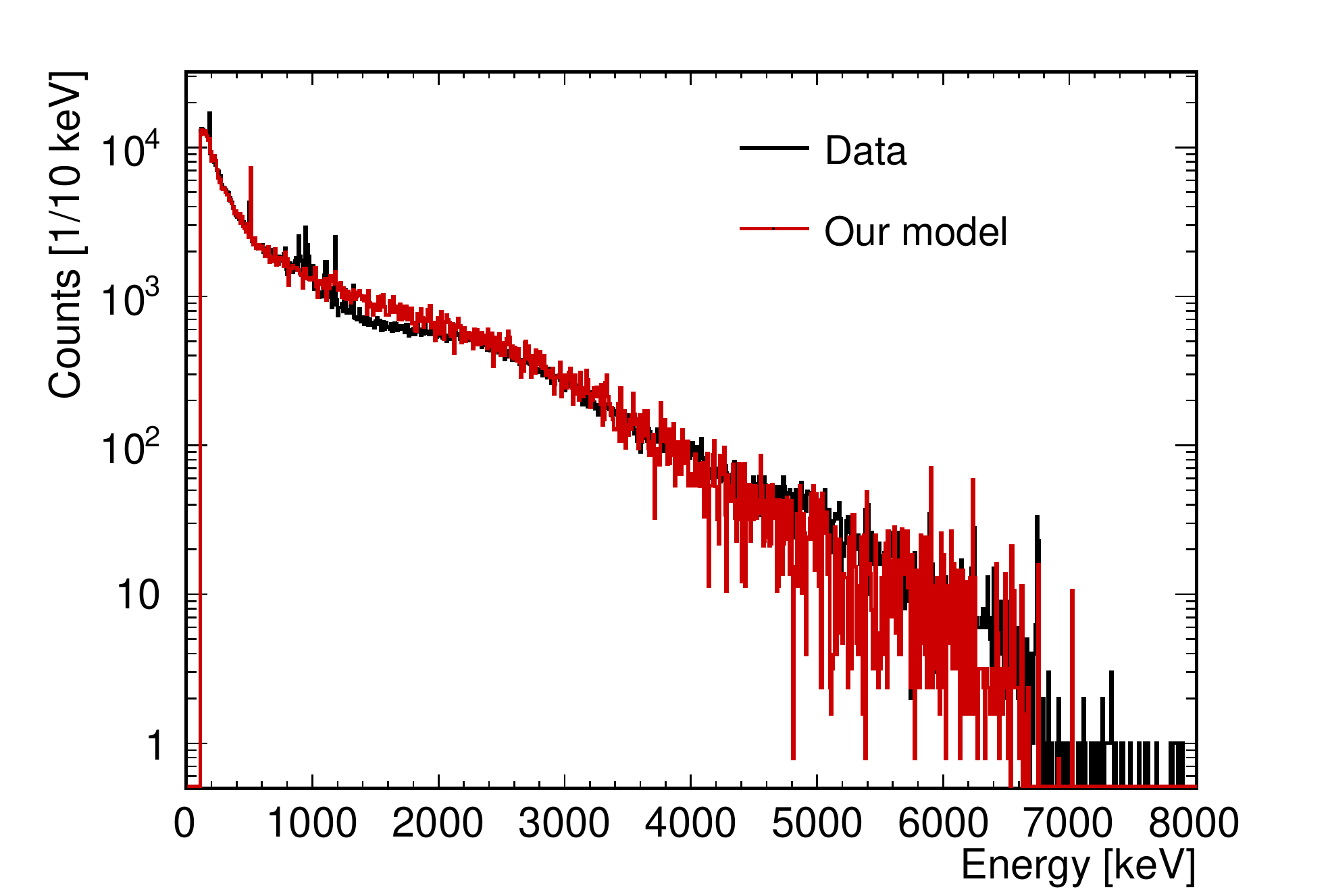}
    \caption{Comparison of the measured energy spectrum from thermal ${}^{157}$Gd$(n,\gamma)$, i.e. data 
             for the peripheral crystal 6 (black) and the energy spectra generated by our model (red): M2H2 (left) and M3H3 (right). }
    \label{fig:mh2233}
  \end{center}
\end{figure}
}
\vspace{2cm}

\bibliographystyle{IMANUM-BIB}
\bibliography{MyBib}

\end{document}